\documentclass[11pt]{article}

\usepackage[T1]{fontenc}
\usepackage{lmodern}
\usepackage[margin=1in]{geometry}
\usepackage[authoryear,round]{natbib}
\usepackage{graphicx}
\usepackage{float}
\usepackage{amsmath,amssymb,amsthm,mathrsfs,mathtools,bm,booktabs,tabularx,microtype,enumitem,xcolor,tikz}
\usetikzlibrary{arrows.meta,positioning,fit,shapes.geometric}
\usepackage[colorlinks,citecolor=blue,urlcolor=blue]{hyperref}
\hypersetup{linkcolor=blue!45!black,citecolor=blue!45!black,urlcolor=blue!55!black}
\theoremstyle{plain}
\newtheorem{theorem}{Theorem}[section]

\newtheorem{proposition}[theorem]{Proposition}
\newtheorem{corollary}[theorem]{Corollary}
\theoremstyle{definition}
\newtheorem{definition}[theorem]{Definition}
\newtheorem{assumption}[theorem]{Assumption}
\theoremstyle{remark}

\newcommand{\PP}{\mathbb P}
\newcommand{\RR}{\mathbb R}

\newcommand{\TV}{d_{\mathrm{TV}}}

\title{Identification and Honest Recovery from Semantic Observation Kernels:
Operator Error, Coarsening, and Stability}
\author{Matthew Dixon\\
\small Artificial Intelligence Finance Institute (AIFI)\\
\small \texttt{matthew.dixon@aifinanceinstitute.com}}
\date{August 2026}

\begin{document}
\maketitle

\begin{abstract}
Probabilistic text generators, such as large language models, assign
probabilities to phrases, but consequential decisions require posterior
uncertainty over meaningful states. These are not interchangeable: language
probabilities depend on the prompt, may be incomplete and need not reliably
identify state uncertainty. Without a statistical bridge, fluent responses
and numerical confidence are insufficient for inference or governance.

We formulate recovery of the target posterior as a semiparametric inverse
problem and develop honest recovery guarantees that account jointly for
calibration error, measurement noise, incomplete probabilities and weak
identification. Simulations demonstrate the predicted coverage and stability
behaviour, while two frozen language-model studies demonstrate held-out
recovery. The resulting method determines when a semantic measurement can be
trusted for inference and when use, review, recalibration or abstention is
warranted, providing a statistical foundation for runtime AI governance.
\end{abstract}


(secondary).

\section{Introduction}

Large language models (LLMs)\footnote{For readers unfamiliar with language
modelling, Table~\ref{tab:llm-terminology} in the appendix translates the few
required terms into their statistical meanings.} assign conditional
probabilities to tokens. The
probability of a complete phrase is obtained by multiplying the probability
of each successive token conditional on the prompt and the preceding tokens.
These probabilities answer
\emph{what language is likely after this prompt?}; scientific use instead asks
\emph{what is the posterior probability of each application-relevant state
given the evidence?} A printed statement such as ``the probability of urgent
intervention is 70\%'' is generated text: it need not equal the probability of
phrases expressing urgency, and neither quantity is automatically a state
posterior. Equivalent rewordings may also redistribute phrase probability
without changing the evidence or intended states.

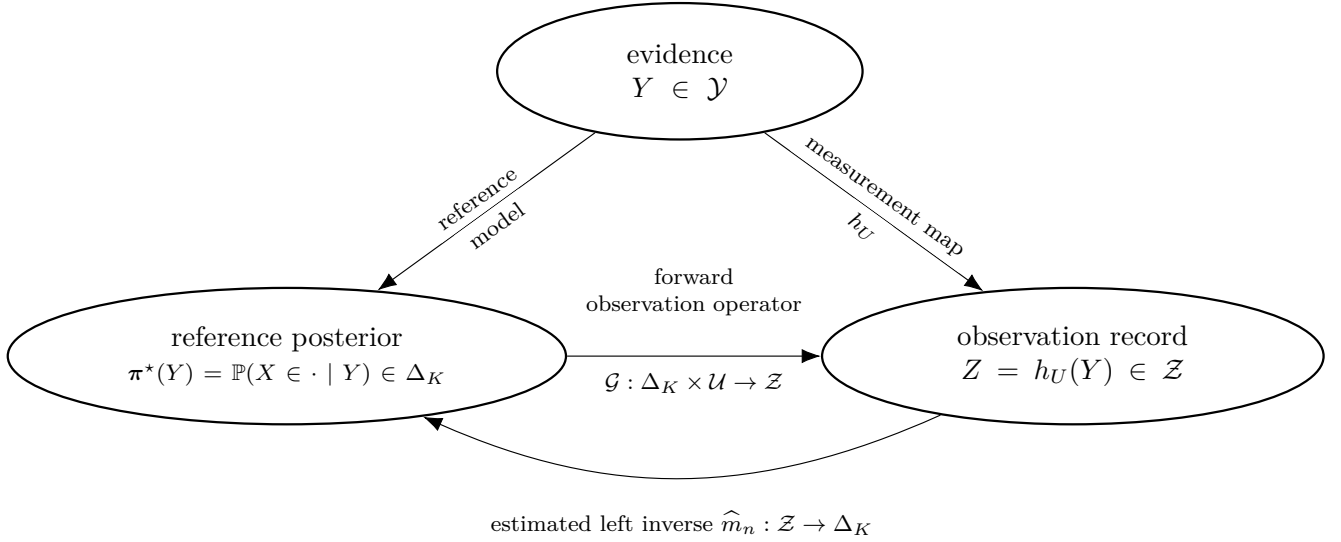
\begin{figure}[t]
\centering
\resizebox{.94\linewidth}{!}{
\begin{tikzpicture}[
  main/.style={ellipse,draw=black,thick,align=center,font=\small,
    minimum height=17mm,inner sep=6pt},
  arr/.style={-{Latex[length=2.4mm]},thin},
  note/.style={font=\scriptsize,align=center,fill=white,inner sep=3pt}
]
\path[use as bounding box] (-7.25,4.15) rectangle (7.25,-2.55);

\node[main,text width=28mm] (evidence) at (0,3.40)
 {evidence\\$Y\in\mathcal Y$};

\node[main,text width=45mm] (target) at (-4.90,-.15)
 {reference posterior\\
 \scriptsize$\bm\pi^\star(Y)
 =\mathbb P(X\in\cdot\mid Y)\in\Delta_K$};

\node[main,text width=40mm] (record) at (4.90,-.15)
 {observation record\\
 $Z=h_U(Y)\in\mathcal Z$};

\draw[arr] (evidence) --
 node[note,above,sloped] {reference}
 node[note,below,sloped] {model} (target);
\draw[arr] (evidence) --
 node[note,above,sloped] {measurement map}
 node[note,below,sloped] {$h_U$} (record);

\draw[arr] (target) --
 node[note,above,yshift=4mm] {forward\\observation operator}
 node[note,below,yshift=-1mm]
 {$\mathcal G:\Delta_K\times\mathcal U\to\mathcal Z$}
 (record);
\draw[arr] (record) to[bend left=24]
 node[note,below,yshift=-3mm]
 {estimated left inverse $\widehat m_n:\mathcal Z\to\Delta_K$}
 (target);

\end{tikzpicture}}
\caption{Model-agnostic inverse-problem formulation. Evidence $Y$ determines
the reference posterior $\bm\pi^\star(Y)=\mathbb P(X\in\cdot\mid Y)$ and,
through a measurement map $h_U$, an observable record $Z=h_U(Y)$. The
forward operator $\mathcal G$ relates the posterior and measurement spaces;
calibration estimates a posterior left inverse. In the LLM application,
$h_u(y)=(u,\bm\pi_u(y))\in\mathcal Z$, where
$\bm\pi_u=\Phi_u\circ\mathcal L_u\circ C_u$. Here $C_u$ constructs the prompt,
$\mathcal L_u(c)=Q_u(\cdot\mid c)$ is the fitted language model's conditional
phrase law, and $\Phi_u$ groups phrase probabilities by declared meaning.}
\label{fig:spaces}
\end{figure}

Statistical calibration is well developed for probabilistic forecasts and
classifiers, through proper scoring rules, general calibration diagnostics
and honest nonparametric calibration bands
\citep{gneiting2007strictly,vaicenavicius2019evaluating,
dimitriadis2023honest}. Comparatively little is known, however, about when a
prompt-dependent language distribution can serve as a statistical measurement
of an independently defined posterior. The surrounding apparatus has largely
been treated as a problem of model engineering or prompt design rather than as
a statistical experiment with a declared estimand, observation law and
identification conditions. Yet probability-valued language output provides an
observable channel from which beliefs over declared states may be recovered,
provided that the channel is defined, calibrated and tested.
Figure~\ref{fig:spaces} shows this conceptual setup as an inverse problem.
Evidence $Y$ defines the prompt-independent reference posterior
$\bm\pi^\star(Y)$ and enters the prompt $C_U(Y)$ under presentation rule $U$.
The LLM supplies phrase
probabilities; semantic grouping produces the prompt-dependent vector
$\bm\pi_U(Y)$; and calibration learns whether the resulting record can be
mapped back to the reference posterior. The simple diagram is the conceptual
arrangement for the statistical setup below. Its purpose is to separate
context dependence from the standard inverse problem,
without allowing language-model terminology common in computer science to
obscure the familiar statistical structure.

This is a measurement problem before it is a prediction problem. Distinct
states may induce the same phrase distribution, returned probabilities may be
truncated, and a good average predictor may still amplify small errors. These
are familiar statistical difficulties in an unfamiliar observation space.
The relevant ideas come from comparison and approximation of
statistical experiments \citep{blackwell1953equivalent,lecam1964sufficiency},
statistical inverse problems
\citep{engl1996regularization,cavalier2008nonparametric}, errors in variables
and generated regressors
\citep{carroll2006measurement,mammen2012generated}, hidden-state
observability \citep{vanhandel2009observability}, and minimax
nonparametrics \citep{tsybakov2009introduction}. The language setting adds
semantic groups---several phrases may represent one state---and
partly observed probability records. Unreturned mass is missing information,
not evidence of zero probability.

\subsection*{A simple example}

We assume that the reader has little or no knowledge of language models. To
build intuition about the problem at hand, consider a triage application with three
declared states: urgent intervention, routine follow-up, and insufficient
information.  A realized evidence record $y$ may contain symptoms,
measurements, and a short
clinical note.  A reference probability model, specified independently of
the language system, assigns
\[
 \bm\pi^\star(y)=(0.60,\,0.30,\,0.10).
\]
This is the target posterior.  It is a probability distribution over
application states, not over words.

The fitted language system operates differently. Two prompts that contain the
same evidence may assign different phrase probabilities:
\begin{center}
\footnotesize
\begin{tabular}{lccccc}
\toprule
\phantom{Prompt} &
\shortstack{urgent\\review} &
\shortstack{immediate\\escalation} &
\shortstack{routine\\follow-up} &
\shortstack{insufficient\\information} &
\shortstack{grouped urgent\\probability}\\
\midrule
Prompt A & .38 & .17 & .35 & .10 & .55\\
Prompt B & .20 & .35 & .35 & .10 & .55\\
\bottomrule
\end{tabular}
\end{center}
No single phrase is the urgent state. The semantic map groups ``urgent
review'' and ``immediate escalation.'' Although rewording moves probability
between the two phrases, both prompts yield the same grouped vector
$(.55,.35,.10)$. This illustrates the desired semantic invariance. It does
not yet make $.55$ a posterior probability. Held-out calibration must still
determine whether grouped vectors of this kind reliably recover the reference
posterior $(.60,.30,.10)$. If rewording instead changed the grouped vector to
$(.35,.55,.10)$, prompt presentation would have altered the measurement and
would need to enter its uncertainty.

The first horizontal arrow aggregates the phrase law over prespecified measurable
semantic cells---sets of meaning-equivalent phrases in the sense defined by
the companion measurement framework \citep{dixon2026calibrating}. Its output remains an observable,
prompt-dependent measurement rather than a posterior.  The second arrow is a
statistical inverse learned from paired semantic measurements and reference
posteriors; the reference posterior is not produced by the language model.
Three questions organize the paper.
Identification asks whether distinct reference posteriors induce distinct
observable language laws. If they do not, no estimator can determine which
posterior generated the observation, regardless of sample size or algorithm.
Stability and learning ask whether the inverse can
be estimated uniformly without excessive amplification of probability error.
Partial identification asks what set of posteriors remains compatible with a
record that exposes only selected phrase probabilities.

\paragraph{Why the standard inverse problem is not enough.}
If the complete semantic probability vector were observed and its forward map
from the reference posterior were known, this would be a standard, possibly
nonlinear inverse problem.  The present setting adds three technicalities.
First, the forward map is not supplied by the language model and must be
learned from paired semantic measurements and reference posteriors.  Second,
the service may return only selected token probabilities, leaving some phrase
and semantic probabilities unobserved; point recovery may then have to be
replaced by a compatible posterior set.  Third, changing the prompt
can change the observation law even when the evidence and target are fixed,
so recovery must be shown to remain stable over the declared prompt and
evidence domain.  The paper's identification, learning and partial-
identification results address these departures from the standard case.

\paragraph{Scope beyond LLMs.}
This is not a paper about natural language as a linguistic object. The
observations are conditional probabilities assigned to prespecified phrases,
calculated from the fitted model's token probabilities. Phrase meanings and
their grouping into scientific states are supplied by the study design rather
than learned by the theory; grammar, discourse and linguistic quality are
outside its scope.
Although LLMs motivate the empirical analysis, the scope of the paper is not
limited to them. More generally, the paper studies recovery of a finite-state
posterior from an indirect, probability-valued observation. The results apply
to survey, diagnostic, expert or sensor probability measurements whenever a
defensible record and reference cases are available. It differs from ordinary
multiclass prediction because the target is the full conditional distribution,
and from conventional classifier calibration because several phrases
may share one meaning while other probabilities are unavailable. The map from
the reference posterior to the reported semantic distribution is therefore an
observation operator, and recovery is an inverse problem. No
transformer-specific quantity enters the theory.

\paragraph{Relation to the companion measurement paper.}
The companion paper \citep{dixon2026calibrating} introduces the semantic
measurement experiment, its deterministic error decomposition and a held-out
validation procedure. Those constructions and results are taken as fixed and
are neither modified nor re-established here. The present paper treats their
output as one example of an indirect probability-valued observation and asks
the downstream statistical question: when does such an observation identify
an independently defined posterior, and with what uncertainty? It develops
the theory needed to answer that question: inverse moduli,
simultaneous stability bands, minimax limits and partial identification under
truncated probability records. Secondary nonparametric learning extensions
are recorded in the appendix. Thus the
companion paper establishes how the measurement is constructed and audited,
whereas this paper characterizes when reliable recovery is statistically
possible and supplies the uncertainty needed to govern its runtime use. In
short, the companion paper constructs the measurement; this paper establishes
when it can be trusted for inference. The logical dependence is one-way for
exposition: the companion
measurement supplies a motivating observation, while the present inverse
theory applies to probability-valued observations beyond language models.

\paragraph{Contribution of the paper.}
After the state definition and observation map have been declared, the paper
answers three questions:
\begin{enumerate}
\item An inverse modulus determines whether the probability-valued experiment
identifies the posterior, quantifies error amplification, and yields
finite-replication and impossibility results.
\item A unified honest-recovery region combines estimated-operator error,
new-record sampling error, coarsening and inverse conditioning. It specializes
to a sharp compatible posterior set under exact truncation and supplies an
explicit diameter guarantee. Here \emph{honest} is used in its standard
frequentist sense: the stated coverage holds uniformly over the declared model
class, rather than only at a selected data-generating law
\citep{gine2010confidence}.
\item Simultaneous derivative bands determine whether stability can be
certified, rejected or left unresolved; a classical minimax boundary shows
that this grey region is unavoidable. Secondary nonparametric learning rates
are recorded in the appendix.
\end{enumerate}
Together, these results determine whether an indirect probability measurement
can be trusted for inference on its declared domain and, when it cannot,
whether the appropriate runtime response is review, recalibration or
abstention. This is the paper's statistical foundation for runtime AI
governance. The simulations demonstrate
inverse-recovery rates, the unified honest region, uniform stability and its
minimax boundary, and coverage of the sharp posterior set under truncation. A
frozen 520-scenario financial market-regime
study supplies exact reference posteriors from a prespecified evidence model
and archived candidate probabilities from two fitted language models.
Filtering, losses and regret require additional transition and decision
models and are outside the present scope.

The individual mathematical ingredients are mostly classical: continuity of
an inverse on a compact domain, minimum-distance arguments, multinomial
concentration, Le Cam's two-point method, finite-difference control of a
derivative, nonparametric derivative lower bounds and sharp identified sets.
The contribution is not to rename these results as language-model theory. It
is to formulate the probability-valued semantic observation experiment, show
which classical result answers each of its failure modes, and connect
observable probability, semantic and missing-mass errors to posterior
identification, stability or set-valued recovery. This connection is absent
from the classical results because they begin with an observation operator and
error metric already specified; constructing and auditing those objects is
precisely the difficulty here.

Table~\ref{tab:theory-lineage} makes this division explicit. A result marked
``classical'' is cited and not proved. Proofs are retained only when the paper
adds a semantic-observation error term, an operational decision rule, or a
sharp set construction specific to truncated probability records.
Theorem~\ref{thm:honest} is the principal new theorem. The
sharp compatible-set and uniform stability results are supporting
paper-specific results; the compact-inverse, Le Cam, concentration and minimax
statements are cited specializations and are not presented as new theory.

\begin{table}[ht]
\caption{Lineage of the theoretical results. ``Specialized'' means that a
classical result is applied to the paper's semantic observation operator; it
does not claim a new general theorem.}
\label{tab:theory-lineage}
\centering\footnotesize
\begin{tabularx}{\linewidth}{p{.22\linewidth}p{.25\linewidth}X}
\toprule
Result & Classical basis & What this paper contributes\\
\midrule
Inverse principle & Compact inverse and minimum distance
\citep{engl1996regularization,tsybakov2009introduction} & Defines the
prompt-dependent semantic law as the observation operator and its inverse
modulus.\\
Plug-in and replication bounds & Error propagation and multinomial
concentration & Separates probability, semantic and observation errors and
propagates them to posterior coordinates.\\
Observational-equivalence bound & Le Cam's two-point method
\citep{lecam1964sufficiency,tsybakov2009introduction} & Makes erased semantic
directions an explicit non-identification failure.\\
Uniform stability gate & Simultaneous bands and finite differences
\citep{gine2010confidence,chernozhukov2014gaussian} & Converts a uniform band
into accept, reject or unresolved decisions for semantic inversion.\\
Minimax boundary & H\"older derivative-estimation lower bounds
\citep{stone1982optimal,tsybakov2009introduction} & Shows that the gate's
unresolved region cannot be removed uniformly.\\
Compatible posterior set & Partial-identification theory
\citep{manski2003partial,tamer2010partial} & Gives the sharp inverse image and
a missing-mass--conditioning diameter bound.\\
Unified honest recovery & Confidence-set inversion and triangle inequalities
& Combines estimated-operator error, new-record noise, set-valued coarsening
and inverse conditioning in one coverage and diameter theorem.\\
\bottomrule
\end{tabularx}
\end{table}

\subsection*{Relation to existing statistical and language-model research}

\paragraph{Statistical experiments, identification, and inverse recovery.}
Figure~\ref{fig:spaces} locates the proposed method, which draws on a
broad array of established statistical literatures. Blackwell comparison and
Le Cam approximation ask what parameter
information an observable experiment retains
\citep{blackwell1953equivalent,lecam1964sufficiency}; inverse-problem theory
asks whether its observation operator is invertible and how it amplifies error
\citep{engl1996regularization,cavalier2008nonparametric}. Here that operator
is induced by a prompt-dependent semantic partition of a conditional language
law and may be only partly observed.

\paragraph{Estimated measurements and semiparametric learning.}
Measurement-error and two-stage semiparametric theory show how a noisy or
estimated covariate changes second-stage convergence and limiting laws
\citep{carroll2006measurement,mammen2012generated}. Grouped language
probabilities are both imperfect measurements and generated regressors, so
our rates separate calibration-sample error from repeated-measurement error.
Related work stresses joint treatment of identification, regularization
and measurement error \citep{emmenegger2021regularizing,shi2021measurement};
our contribution is the decomposition for a partly observed verbal law.

\paragraph{Partial identification.}
Omitted probability mass calls for an identified set, not zeros assigned to
unreturned responses. Following partial-identification and set-valued
inference \citep{manski2003partial,tamer2010partial,
chernozhukov2007estimation,beresteanu2011randomsets}, we construct the sharp
simplex-valued set, carry it through the inverse, and bound its diameter by
missing mass and inverse conditioning.

\paragraph{Compositions, calibration, and predictive coverage.}
Because target and observed probabilities are compositions, calibration uses
log-ratio coordinates \citep{aitchison1982statistical,egozcue2003isometric}
and proper scores \citep{gneiting2007strictly}. Conformal methods quantify
held-out recovery error \citep{lei2018distributionfree,
barber2021jackknife,angelopoulos2023conformal}, subject to their sampling
assumptions \citep{barber2023beyond}; coverage does not itself establish
identification or transport to a new prompt or evidence domain.

\paragraph{Language-model uncertainty.}
Language-model work documents miscalibration across token, sequence and answer
levels \citep{braverman2020calibration,guo2017calibration,
jiang2021calibration}; elicited confidence can differ from native
probabilities \citep{tian2023justask,giovannotti2024calibrated}. Semantic
entropy groups sampled answers by meaning
\citep{kuhn2023semantic,farquhar2024semantic}, while decision-theoretic work
emphasizes task dependence \citep{wang2025subjective}. We instead seek a full
probability distribution over prespecified states and ask whether the
observable experiment identifies it. Prompt variation remains in the design,
unexpressed mass remains visible, and an independent posterior defines
recovery. This complements broader surveys of language-model uncertainty
\citep{xia2025survey} with explicit inverse and minimax conditions.

\paragraph{Uniform stability as a statistical functional.}
Uniform inverse control is stronger than accurate prediction at sampled
points. Results on nonlinear filtering control the sensitivity or forgetting
of a specified recursive filter as observations accumulate
\citep{legland2000forgetting,vanhandel2009uniform}. They do not establish that
an unknown observation inverse learned from calibration data is uniformly
well conditioned, nor do they provide a simultaneous confidence statement for
its sensitivity over the supported domain. The derivative functional studied
here addresses that distinct question: can worst-case local sensitivity be
bounded with stated coverage wherever the inverse will be used? The result is
entirely an external statistical measurement, not a claim about a language
model's internal belief.

\subsection*{Overview}

The central statistical problem this paper addresses is whether an indirect,
prompt-dependent probability measurement permits reliable recovery of a
posterior over meaningful states.
The paper proceeds from the weakest claim to the strongest. Section~2 starts
with the canonical inverse problem, then addresses what makes the language
setting nonstandard: the numerical observation must be constructed by grouping
phrase probabilities by meaning, its forward map must be learned, and the
service may reveal only a coarsened record. The section determines whether
this learned, partially observed probability channel identifies the target and
uses an inverse modulus to quantify whether recovery is stable enough for
estimation.
Section~3 turns that population model into finite-sample inference. It defines
the calibration estimator, constructs a stability certificate, and treats
incomplete records through sharp set-valued inference. Secondary uniform
learning theory is collected in the appendix because it is not exercised
by the present experiments. Section~4 specifies the theory-guided simulations
and frozen language study. Section~5 reports the results and discusses exactly
what they support and what remains assumption-dependent. Section~6 draws the findings together into practical
conditions for auditable state measurement.
Proofs are collected in the appendix.

\section{Statistical model and population identification}

We begin with the simple inverse problem in Figure~\ref{fig:spaces}. Evidence
$Y$ determines an independently defined target posterior
$\bm\pi^\star(Y)$. A measurement channel turns the same evidence into an
observable record $Z=h_U(Y)$. The forward operator relates the posterior to
that record; calibration estimates a left inverse from $Z$ to the posterior
simplex. Everything else in this section---phrase laws, semantic grouping,
coarsening and log-ratio coordinates---makes one of these four objects
precise. None changes the estimand.

\subsection{Probability experiment, target and observation}

Let $(\Omega,\mathscr F,\mathbb P)$ be the probability space for the
calibration experiment. It supports a finite scientific state
$X\in\mathcal X=\{x_1,\ldots,x_K\}$, evidence $Y\in\mathcal Y$, and a
presentation rule $U\in\mathcal U$.  The scientific target is defined
independently of the language model:
\[
 \bm\pi^\star(y)
 =\bigl(\mathbb P\{X=x_k\mid Y=y\}\bigr)_{k=1}^K\in\Delta_K,
 \qquad
 \Delta_K=\left\{\bm p\in[0,1]^K:\sum_{k=1}^Kp_k=1\right\}.
\]
Thus $\bm\pi^\star(Y)$ is random through $Y$, but it is unchanged by the
presentation rule. The same evidence also produces the observable record
\[
 Z=h_U(Y)\in\mathcal Z.
\]
Here $h_u:\mathcal Y\to\mathcal Z$ is the complete observation map under
rule $u$. In the language-model application,
\[
 h_u(y)=(u,\bm\pi_u(y)),
 \qquad
 \bm\pi_u=\Phi_u\circ\mathcal L_u\circ C_u,
\]
where $C_u$ constructs the prompt, $\mathcal L_u$ returns the fitted model's
conditional phrase law, and $\Phi_u$ groups phrase probabilities by declared
meaning. These components are defined below. Recording $u$ identifies the
measurement channel; it does not make the prompt part of the posterior target.

\subsection{Why this is an inverse problem}

The canonical pointwise formulation assumes that the observation depends on
evidence through its posterior and the recorded presentation rule:
\[
 h_u(y)=\mathcal G\{\bm\pi^\star(y),u\},
 \qquad
 \mathcal G:\Delta_K\times\mathcal U\longrightarrow\mathcal Z.
\]
The operator $\mathcal G$ describes what the indirect measurement retains
about the target. A recovery rule $m_0:\mathcal Z\to\Delta_K$ is a posterior
left inverse when
\[
 m_0\{\mathcal G(\bm\pi,u)\}=\bm\pi
\]
on the supported domain. The calibration sample is simply
\[
 \mathcal D_n
 =\{(z_i,\bm\pi^\star(y_i)):z_i=h_{u_i}(y_i),\ i=1,\ldots,n\}.
\]
Calibration fits $\widehat m_n$ from these paired indirect measurements and
independently supplied targets. A critical departure from ordinary posterior
regression is that $\widehat m_n$ does not receive $y_i$ as an input. The
evidence, presentation rule and complete prompt remain in the audit record for
provenance and held-out design checks, but supplying $y_i$ to the fitted map
would allow it to predict the target directly. That would bypass the
measurement channel and would no longer test whether the language-derived
observation identifies the posterior.

This exposes the two population questions. Identification asks whether two
targets can produce the same observation. Stability asks how much the inverse
magnifies a small observational error. A prediction score alone answers
neither question.

\paragraph{A secondary complication: a law-valued forward operator.}
The pointwise factorisation is the cleanest case, but it need not hold. Two
evidence records can have the same posterior and still produce different
observations. The general forward experiment then uses the conditional law
\[
 \mathsf P_{\bm p}^{O}
 =\mathcal L_{\mathbb P}\{Z^{O}\mid
   \bm\pi^\star(Y)=\bm p\},
 \qquad
 T^O(\bm p)=\mathsf P_{\bm p}^{O},
\]
where $O$ is the coarsening rule introduced below. The simple $\mathcal G$ is
the deterministic special case. The law-valued $T^O$ accommodates repeated
measurements, heterogeneous evidence and incomplete records; identification
and stability are then questions about observable laws rather than single
vectors.

\subsection{How phrase probabilities become a measurement}

Let $\mathcal U$ be a finite family of presentation rules. A presentation
rule is a prespecified deterministic template that fixes the task instruction
and the placement, ordering and format of the evidence. For example, one rule
may present ``temperature $39.4^\circ$C; oxygen saturation $89\%$'' before the
triage question, while another asks the same question first and then supplies
the same two measurements in reverse order. For $u\in\mathcal U$, the
measurable map
$C_u:(\mathcal Y,\mathscr Y)\to(\mathcal C,\mathscr C)$ inserts evidence into
rule $u$ and returns the complete prompt or context.  Let
$(\mathcal R,\mathscr R)$ be the measurable space of complete phrases.
The fitted language system supplies the otherwise
unrestricted conditional kernel
\[
 Q_u:\mathcal C\times\mathscr R\longrightarrow[0,1],
 \qquad (c,A)\longmapsto Q_u(A\mid c).
\]
Here $Q_u(A\mid c)$ is the probability assigned by the fixed fitted system
that the generated phrase belongs to the measurable set $A\in\mathscr R$ when
the supplied context is $c\in\mathcal C$. In particular,
$Q_u\{A\mid C_u(y)\}$ is that probability for evidence $y$ presented under
rule $u$.
This component is nonparametric.  A declared measurable semantic map
$\phi_u:\mathcal R\to\{0,1,\ldots,J-1\}$ groups meaning-equivalent
phrases. Each inverse image $\phi_u^{-1}\{j\}$ is called a semantic cell: it
is simply the measurable set of phrases assigned to category $j$. Residual
category $0$ records language outside the declared meanings. It is not a null
scientific state: if a null state is part of the application, it is included
among $x_1,\ldots,x_K$ and receives a coordinate in the target posterior.
The corresponding grouping operator is
\[
 \Phi_u\{Q_u(\cdot\mid c)\}
 =\bigl(Q_u\{\phi_u^{-1}(j)\mid c\}\bigr)_{j=0}^{J-1}.
\]
Consequently $\bm\pi_u(y)=(\Phi_u\circ\mathcal L_u\circ C_u)(y)$ is the
observable semantic probability vector appearing in Figure~\ref{fig:spaces};
it is a measurement of, not yet an estimate equal to, the reference posterior.

For example, fix a prompt $c$ and suppose the phrases ``urgent review'' and
``immediate escalation'' are both mapped to the urgent state. Thus
\[
 A=\phi_u^{-1}\{\mathrm{urgent}\}
 =\{\text{``urgent review''},\text{``immediate escalation''}\}
 \in\mathscr R.
\]
If their probabilities are $.38$ and $.17$, respectively, then
\[
 Q_u(A\mid c)=.38+.17=.55.
\]
A phrase outside every declared meaning is instead assigned to residual
category $0$.

\subsection{Coordinate representation}

The conceptual target remains the posterior $\bm\pi^\star(y)$ on the
simplex. Euclidean coordinates are introduced only because nonparametric
estimation, derivatives and inverse moduli are more naturally stated on a
subset of $\RR^q$ than on a constrained simplex. Let
\[
 \operatorname{int}(\Delta_K)
 =\{\bm p\in\Delta_K:p_k>0\text{ for every }k\},
\]
let $\Pi_0\subset\operatorname{int}(\Delta_K)$ be the supported posterior
domain, and choose a prespecified continuous one-to-one coordinate map
\[
 \boldsymbol\ell:\Pi_0\longrightarrow\Xi\subset\RR^q,
 \qquad
 \bm\xi(y)=\boldsymbol\ell\{\bm\pi^\star(y)\}.
\]
For an unrestricted $K$-state posterior, an orthonormal log-ratio map gives
$q=K-1$ and removes the unit-sum constraint while respecting the relative
geometry of compositions. The inverse $\boldsymbol\ell^{-1}$ returns every
estimate and uncertainty region to probability scale. Thus log-ratio
coordinates solve a geometric problem; they do not redefine the estimand or
alter Figure~\ref{fig:spaces}. A lower-dimensional $q$ is possible when the
scientific model restricts the supported posterior family.

The symbol $J$ below counts observable semantic categories and need not equal
$K$ or $q+1$. For example, adding a residual category gives $J=K+1$, while
the unrestricted posterior still has $q=K-1$ coordinates.

\subsection{What is actually observed?}

The inverse theory is indexed by the posterior coordinate $\bm\xi$, rather
than by an individual evidence record. Several records can have the same
reference posterior coordinate while producing different phrase laws. To
obtain one well-defined observation law at each $\bm\xi$, let
$\overline Q_u(\cdot\mid\bm\xi)$ denote a regular conditional version of
the phrase law induced by the calibration design conditional on
$\bm\xi(Y)=\bm\xi$.  Equivalently, it averages
$Q_u\{\cdot\mid C_u(Y)\}$ over evidence records having that target coordinate;
when the design contains one such record, no averaging is involved. In plain
English, this gives the forward operator one value at each $\bm\xi$, rather
than a value that depends on an arbitrary choice among evidence records with
the same target. It makes the forward operator well defined; it does not make
it identifiable. Identification still requires different target coordinates
to induce different observable laws, and averaging may reduce that
separation. The
semantic observation surface is the map
\[
 \bm g_u:\Xi\longrightarrow\Delta_J\subset\mathbb R^J,
 \qquad
 \bm\xi\longmapsto\bm g_u(\bm\xi)
 =\bigl(\overline Q_u\{\phi_u^{-1}(j)\mid\bm\xi\}\bigr)_{j=0}^{J-1}
 .
\]
Thus $\bm g_u(\bm\xi)$ is a $J$-component probability vector, with one
coordinate for each semantic category $j=0,\ldots,J-1$; geometrically it lies
in the $(J-1)$-dimensional simplex $\Delta_J$. The number $J$ describes the
observable categories and should not be confused with $q$, the dimension of
the target-posterior coordinates.
A finite-dimensional calibration map from these probabilities to
$\bm\xi$, and hence back to $\bm\pi^\star$, completes the semiparametric
model.

Continuing the example, suppose two evidence records $y_1$ and $y_2$ have the
same reference posterior
\[
 \bm\pi^\star(y_1)=\bm\pi^\star(y_2)=(.60,.30,.10),
 \qquad
 \bm\xi(y_1)=\bm\xi(y_2)=\bm\xi,
\]
but, under the same presentation rule, yield grouped phrase vectors
\[
 (0,.55,.35,.10)
 \quad\text{and}\quad
 (0,.59,.31,.10),
\]
where the coordinates represent residual category $0$, urgent, routine and
insufficient information. If the two records receive equal conditional weight
in the calibration design, $\overline Q_u(\cdot\mid\bm\xi)$ averages their
phrase laws and therefore
\[
 \bm g_u(\bm\xi)=(0,.57,.33,.10).
\]
Thus the forward surface uses $(0,.57,.33,.10)$ rather than selecting either
record arbitrarily. In a discrete calibration design, if only one record has
coordinate $\bm\xi$, $\overline Q_u$ reduces to its phrase law.
Calibration then learns from paired calibration cases the relation between the observable input
$\bm g_u(\bm\xi)=(0,.57,.33,.10)$ and the target
$\bm\pi^\star=(.60,.30,.10)$; held-out cases assess whether that relation
generalises. The present paper concentrates on
identification, learning rates,
stability certification and partial identification of this inverse.

The service may reveal only part of this vector. Represent that restriction by
the observation rule
\[
 O:(\Delta_J,\mathcal B(\Delta_J))
 \longrightarrow(\mathcal Z,\mathscr Z),
\]
where $\mathcal Z$ is the space of returned records. When $O$ is the identity,
the full semantic vector is observed. Otherwise, different full vectors can
produce the same record. For example, if
$\bm g_u(\bm\xi)=(0,.57,.33,.10)$ and only the two largest coordinates are
returned, $O$ records $.57$ and $.33$, their total mass $.90$, and that the
remaining $.10$ is unobserved.

Let $U$ denote the recorded presentation rule and $\bm S$ the returned record.
The observable law at target $\bm\xi$ is simply
\[
 \mathsf P_{\bm\xi}^{O}
 =\mathcal L_{\mathbb P}\{(U,\bm S)\mid\bm\Xi=\bm\xi\},
 \qquad
 \mathcal E^{O}=\{\mathsf P_{\bm\xi}^{O}:\bm\xi\in\Xi\}.
\]
Here $\mathcal L_{\mathbb P}$ denotes probability law under the calibration
design. Thus the forward operator is
$T^O(\bm\xi)=\mathsf P_{\bm\xi}^{O}$: target coordinate in, distribution of
observable records out. Recording $U$ allows several prespecified prompt rules
to be used without treating prompt choice as part of the posterior target.

\subsection{When does the observation identify the target?}

We can now state the first question precisely.  If two posterior coordinates
produce the same observable law, no amount of repeated sampling can tell them
apart.  If nearby observable laws can correspond to very distant posterior
coordinates, recovery is unique in principle but unstable in practice.  The
following definition formalizes this distinction through identification and
the inverse modulus.

\begin{definition}[Identification and inverse modulus;
\citealp{engl1996regularization,tsybakov2009introduction}]
The posterior coordinate is identified on $D\subseteq\Xi$ when
$\mathsf P_{\bm\xi}^{O}=\mathsf P_{\bm\xi'}^{O}$ implies
$\bm\xi=\bm\xi'$.  For a metric $d_{\mathcal E}$ on observable laws,
the inverse modulus is
\[
 \omega_D(\varepsilon)
 =\sup\{\|\bm\xi-\bm\xi'\|:
 d_{\mathcal E}(\mathsf P_{\bm\xi}^{O},
 \mathsf P_{\bm\xi'}^{O})\le\varepsilon\}.
\]
\end{definition}

The distinction between identification and conditioning is essential.
Identification is $\omega_D(0)=0$; stable identification additionally
requires $\omega_D(\varepsilon)\downarrow0$.  A one-to-one but nearly flat
semantic response surface can satisfy the first condition and still amplify a
small probability error into an unusable posterior error.

\paragraph{Worked binary example.}
To gain further intuition, we return to the triage setting and let
$\xi\in[-2,2]$ be the reference
log-odds of the urgent state.  Suppose a fixed prompt and semantic grouping
produce an urgent phrase probability
$g_a(\xi)=\{1+\exp(-a\xi)\}^{-1}$.  The inverse is
$g_a^{-1}(p)=a^{-1}\log\{p/(1-p)\}$.  When $a=1$, changing the observed
grouped phrase probability from $.60$ to $.61$ changes the recovered coordinate from
$0.405$ to $0.447$, a difference of about $.042$.  When $a=.05$, the same
probability perturbation changes the coordinate by about $.84$.  Both maps
are one-to-one, but under the second language channel the phrase probabilities
respond only weakly to the clinical state and inversion is twenty times more
sensitive. In the notation below, using absolute distance and
$\varepsilon=.01$, these two pairs imply
$\omega_D(.01)\ge.042$ when $a=1$ but
$\omega_D(.01)\ge.84$ when $a=.05$. Thus the inverse modulus records directly
that the same $.01$ observation error can produce twenty times as much error
in the recovered coordinate.

The next theorem packages two classical facts for this observation experiment:
a continuous injection from a compact domain has a continuous inverse on its
image, and a minimum-distance estimator inherits an inverse-modulus error
bound \citep{engl1996regularization,tsybakov2009introduction}. Its novelty is
therefore not the topological inequality itself, but the identification of the
observable semantic law as the operator to which it must be applied.

\begin{proposition}[Compact-inverse principle;
\citealp{engl1996regularization,tsybakov2009introduction}]
\label{thm:identification}
Suppose $D$ is compact and
$\bm\xi\mapsto\mathsf P_{\bm\xi}^{O}$ is continuous in
$d_{\mathcal E}$.  The inverse on the image exists and is continuous if and
only if the observation map is injective.  If an approximate minimum-distance
estimator satisfies
\[
 d_{\mathcal E}(\widehat{\mathsf P},
 \mathsf P_{\widehat{\bm\xi}}^{O})
 \le\inf_{\bm v\in D}
 d_{\mathcal E}(\widehat{\mathsf P},\mathsf P_{\bm v}^{O})+\eta,
 \quad
 d_{\mathcal E}(\widehat{\mathsf P},
 \mathsf P_{\bm\xi}^{O})\le\varepsilon,
\]
then
\[
 \|\widehat{\bm\xi}-\bm\xi\|
 \le\omega_D(2\varepsilon+\eta).
\]
\end{proposition}

This is a direct specialization of the compact-inverse and
minimum-distance principles cited above; no new general inverse theorem is
claimed. In plain language, Proposition~\ref{thm:identification} separates two questions
that predictive validation often conflates.  Injectivity answers whether
there is a unique answer in the population.  The shape of $\omega_D$ answers
whether finite probability error permits a useful answer.  A low training
loss cannot replace either condition. Conversely, the proposition does not
require a parametric model for language: it applies to the observable law as
a black-box stochastic kernel.

\begin{corollary}[Interface from measurement error to inverse recovery]
\label{cor:plugin}
Let $\bm g(\bm\xi)$ be the ideal semantic observation law and let
$\widehat{\bm g}(\bm\xi)$ be obtained from an estimated phrase
kernel, an estimated semantic grouping, and an observation operator.  Suppose,
uniformly on $D$,
\[
 \|\widehat{\bm g}(\bm\xi)-\bm g(\bm\xi)\|_1
 \le
 \varepsilon_{\rm prob}+\varepsilon_{\rm sem}
 +\varepsilon_{\rm obs}
 =:\varepsilon_{\rm tot},
\]
where the three terms bound, respectively, probability calculation,
semantic-cell mismatch, and incomplete or perturbed observation.  If
$d_{\mathcal E}(\widehat{\mathsf P},\mathsf P_{\bm\xi}^{O})
\le L_O\varepsilon_{\rm tot}$ and the minimum-distance tolerance is $\eta$,
then
\[
 \|\widehat{\bm\xi}-\bm\xi\|
 \le
 \omega_D\!\left(2L_O\varepsilon_{\rm tot}+\eta\right).
\]
When $\omega_D(t)\le C_Gt$ locally, each upstream component is amplified by
at most $2C_GL_O$ in posterior coordinates.
\end{corollary}

The upstream measurement analysis, including the companion paper, supplies or
bounds the three components of $\varepsilon_{\rm tot}$; the present paper does
not rederive that decomposition. Corollary~\ref{cor:plugin} is the interface
that carries the resulting disturbance radius through the inverse modulus.
Its role is therefore to quantify inverse uncertainty and
weak-identification amplification, not to reconstruct the semantic map.

Having established what observation error does to the recovered state, the
next step is to determine how observation error decreases with repeated
measurements.  For a finite semantic partition, ordinary multinomial
concentration supplies that link. The square-root rate is classical; the
corollary propagates it through the semantic inverse rather than claiming a
new concentration inequality.

\begin{corollary}[Finite replication]
\label{cor:replication}
Suppose a single prompt has a finite number $J$ of semantic cells, the full cell vector
$\bm g(\bm\xi)$ is observed through $R$ conditionally independent
multinomial draws, and total variation is used in
Proposition~\ref{thm:identification}. With probability at least $1-\alpha$,
\[
 \|\widehat{\bm\xi}_R-\bm\xi\|
 \le
 \omega_D\!\left[
 \sqrt{\frac{2\{J\log 2+\log(1/\alpha)\}}{R}}+\eta_R
 \right].
\]
If $\omega_D(t)\le C_Gt$ locally and $\eta_R=o(R^{-1/2})$, then
$\widehat{\bm\xi}_R-\bm\xi=O_{\PP}(R^{-1/2})$.
\end{corollary}

\paragraph{Worked replication example.}
For a fixed evidence--prompt pair, suppose the grouped phrase law over
urgent, routine and insufficient-information phrases is $(.55,.35,.10)$.
If $R=100$ independently sampled phrases yield counts $(52,38,10)$,
the empirical semantic law is $(.52,.38,.10)$ and its total
variation distance from the target law is $.03$.  If the local inverse
constant is $C_G=2$, the resulting first-order state error is approximately
bounded by $.06$, before optimization error.  Increasing repetitions from
$100$ to $400$ halves the characteristic sampling error; it does not remove
bias from assigning a phrase to the wrong state or from changing the prompt.

Corollary~\ref{cor:replication} therefore gives a finite-sample statement and
the familiar square-root rate tested below. A local asymptotic normal law for
smooth full-rank measurements is recorded in Appendix~\ref{app:secondary}; it
is not needed for the empirical argument.

The positive results also reveal a sharp failure case.  If two different
targets generate statistically indistinguishable observations, regularizing
or collecting more of the same data cannot identify which target is present.
The following is Le Cam's classical two-point lower bound specialized to the
semantic observation experiment \citep{lecam1964sufficiency,tsybakov2009introduction}.

\begin{proposition}[Le Cam lower bound;
\citealp{lecam1964sufficiency,tsybakov2009introduction}]
\label{thm:impossibility}
Let two admissible systems induce sample laws
$\mathbb Q_0^{(n)},\mathbb Q_1^{(n)}$ but target coordinates separated by
$\Delta=\|\bm\xi_1-\bm\xi_0\|$.  Then
\[
 \inf_{\widehat{\bm\xi}}\sup_{j\in\{0,1\}}
 \mathbb E_{\mathbb Q_j^{(n)}}[\|\widehat{\bm\xi}-\bm\xi_j\|]
 \ge \frac{\Delta}{4}
 \{1-\TV(\mathbb Q_0^{(n)},\mathbb Q_1^{(n)})\}.
\]
In particular, if the observable laws coincide, no estimator is uniformly
consistent.
\end{proposition}

\paragraph{Worked equivalence example.}
Suppose $\xi$ is a signed reference log-odds, with positive values denoting
urgent rather than routine evidence, but the grouped phrase law satisfies
$g(\xi)=g(-\xi)$ because it responds only to $\xi^2$.  Then
$\xi=.8$ and $\xi=-.8$ produce the same phrase probabilities even
though they represent opposite scientific conclusions and are $1.6$ units
apart. Proposition~\ref{thm:impossibility} implies a non-vanishing worst-case
error. This is not an
algorithmic defect: the language observation has erased the direction.
Restoring it requires a more informative prompt, semantic partition, or
observation---not a more elaborate inverse fitted to the same law.

\paragraph{Answer to Question 1.}
A semantic inverse can exist,
be unique, and still be poorly conditioned; or it can fail to exist because
the observable experiment merges distinct targets.  The next section assumes
that a supported inverse is meaningful and studies how it can be learned and
certified from calibration data.

\section{Estimation, uncertainty and partial identification}

\paragraph{Question 2: can the inverse be learned and certified as stable?}
Identification is a population property. In practice the inverse is estimated
from finitely many calibration cases, and low average prediction error does
not rule out severe amplification at some supported inputs. This section asks
how quickly the whole inverse surface can be learned and whether its worst
local sensitivity can be bounded with simultaneous uncertainty.

\subsection{Calibration data and the predictive inverse}

There are two sampling layers. Across calibration scenarios, the semantic probability
summary varies with the evidence.  Within a scenario, that summary may itself
be measured with error because only finitely many repeated phrases or
returned alternatives are available.  For example, returned urgent phrases
may total $.52$ while unreturned alternatives retain $.08$ of unknown semantic
mass.  The theory must account for both sources of uncertainty.

As defined in Section~2, a calibration case is the pair
$(\bm S_i,\bm\Xi_i)$ of an observed semantic-probability record and the
coordinate of its independently specified reference posterior.  Assume
\[
 (\bm S_i,\bm\Xi_i)\stackrel{\mathrm{i.i.d.}}{\sim}
 P_{\bm S,\bm\Xi},\qquad i=1,\ldots,n.
\]
This joint law across calibration cases is distinct from
$\overline Q_u(\cdot\mid\bm\xi)$, the conditional phrase law at a fixed
target and presentation rule. Define
\[
 m_0(\bm S_i)
 =\mathbb E_{P_{\bm S,\bm\Xi}}[\bm\Xi_i\mid\sigma(\bm S_i)],
 \qquad
 m_0(\bm s)
 =\mathbb E_{P_{\bm S,\bm\Xi}}[\bm\Xi_i\mid\bm S_i=\bm s].
\]
This conditional mean is the
identified predictive inverse; it equals a structural inverse only when
$\bm\Xi_i=m_0(\bm S_i)$ almost surely.  This distinction prevents
calibration accuracy from being misreported as structural identification.

\paragraph{Predictive versus structural inversion.}\leavevmode\par
Suppose two clinical records, presented through the same prompt, both yield
the grouped phrase vector $\bm S=(.55,.35,.10)$, but their independently
specified reference coordinates are $.40$ and $.50$ because the language
measurement omits a relevant aspect of the evidence.  The predictive inverse is their
conditional mean, approximately $.45$.  It can be estimated and predict well,
but it is not a one-to-one structural inverse: $\bm S$ does not determine the
reference coordinate exactly.  This distinction is why the paper writes
$m_0(\bm s)=\mathbb E_{P_{\bm S,\bm\Xi}}[\bm\Xi\mid\bm S=\bm s]$ before imposing stronger
conditions.

\paragraph{Secondary learning theory.}
Uniform nonparametric learning rates are useful when the inverse surface is
estimated flexibly, but the present experiments do not test those rates
directly. To keep the empirical argument focused, Appendix~\ref{app:secondary}
states the independent-sample rate, its generated-regressor contribution and
a geometrically beta-mixing extension. The main text proceeds with the
stability statement actually assessed below. All conclusions remain local to
the frozen prompt, evidence and calibration support; held-out agreement cannot
establish global injectivity or transport to an untested domain.

\subsection{Certifying stability}

Why introduce a derivative at all? Small mean prediction error can coexist
with severe local amplification of measurement noise. The derivative records
that sensitivity, and its supremum supplies a uniform rather than average
certificate. Let $F_0(\bm z,\bm w)$ be the conditional-mean calibration
surface for the recovered coordinate and define
\[
 \kappa(F_0)=
 \sup_{(\bm z,\bm w)\in D_0}
\|D_{\bm z}F_0(\bm z,\bm w)\|_{\mathrm{op}}.
\]

\paragraph{Sensitivity intuition.}
Let $z$ be the log-ratio of grouped urgent to routine phrase
probabilities under a fixed prompt.  Two fitted calibrations,
$F(z)=0.8z$ and $F(z)=1.05z$, may have similar mean posterior error on a
concentrated test sample.  Yet a $.10$ change in the language log-ratio is
locally transmitted as at most $.08$ by the first map and $.105$ by the
second. The relevant distinction is the derivative, not merely the fitted
level. A statistically valid gate must therefore cover the derivative
uniformly, including its worst supported point.

\begin{theorem}[Uniform stability gate]\label{thm:band}
Suppose $D_{\bm z}F_0$ is Lipschitz and a simultaneous band satisfies
$\|\widehat F-F_0\|_{\infty,D_0}\le r_n(\alpha)$ with probability at least
$1-\alpha$.  A centered finite-difference Jacobian with step $h$ yields
\[
 |\widehat\kappa-\kappa(F_0)|
 \le C_1h+C_2r_n(\alpha)/h.
\]
Taking $h\asymp r_n(\alpha)^{1/2}$ gives a radius
$\rho_n(\alpha)=O\{r_n(\alpha)^{1/2}\}$.  Therefore
\begin{align*}
 \widehat\kappa+\rho_n(\alpha)<1
 &\quad\Longrightarrow\quad \text{sensitivity below the unit threshold},\\
 \widehat\kappa-\rho_n(\alpha)>1
 &\quad\Longrightarrow\quad \text{sensitivity above the unit threshold}.
\end{align*}
All remaining cases are statistically unresolved.
\end{theorem}

Theorem~\ref{thm:band} deliberately permits a third decision:
``unresolved.'' If the confidence interval for $\kappa(F_0)$ crosses the
prespecified unit threshold, the data support neither side. The unit value is
a normalization; an application may prespecify another acceptable
sensitivity after placing input and output in commensurate coordinates.

The generic result starts from a simultaneous band for the function and
converts it into a derivative bound using the classical bias--stochastic-error
trade-off for finite differences. Its contribution is the resulting
three-way operational gate for the worst supported sensitivity, not a new
finite-difference inequality. When the
update is known to lie in a finite-dimensional Gaussian sieve, a sharper
finite-sample construction is possible. Corollary~\ref{cor:gaussianband} in
the appendix gives the exact band used in the simulation. It is secondary to
the main argument because it assumes a correctly specified Gaussian sieve;
Theorem~\ref{thm:band} is the general stability result. The important point is
simultaneity: separate pointwise intervals do not cover the worst derivative
over the whole supported domain.

Finally, we ask whether the slow derivative rate is merely an artifact of
the proposed estimator. Proposition~\ref{thm:minimax} is a specialization of the
classical minimax lower-bound construction for estimating derivatives of a
H\"older regression function \citep{stone1982optimal,tsybakov2009introduction}.
The rate itself is therefore not claimed as new. Its role here is to show that
the unresolved region of the semantic stability gate is statistically
unavoidable. This emphasis on simultaneous validity and minimax resolution is
also shared by modern work on distributional functionals
\citep{manole2022minimax}.

\begin{proposition}[Minimax derivative boundary;
\citealp{stone1982optimal,tsybakov2009introduction}]\label{thm:minimax}
Let $D_0$ contain a $d$-dimensional interior ball and let $F$ range over a
H\"older ball of smoothness $s>1$.  Under a design density bounded above and
below and Gaussian regression noise, let $\mathbb P_F$ denote the resulting
data law. Then
\[
 \inf_{\widehat\kappa}\sup_F
 \mathbb E_{\mathbb P_F}[|\widehat\kappa-\kappa(F)|]
 \ge c\,n^{-(s-1)/(2s+d)}.
\]
No test uniformly distinguishes $\kappa(F)<1$ from $\kappa(F)>1$ when the
separation from one is $o\{n^{-(s-1)/(2s+d)}\}$.
\end{proposition}

Proposition~\ref{thm:minimax} explains why low held-out prediction error cannot,
by itself, establish uniform local stability. Stability is a derivative functional
and has a slower resolution scale than function prediction.

\paragraph{Practical consequence.}
If the estimated worst-case response of posterior coordinates to a change in
the language-probability coordinates is $.98$ with a valid radius $.05$, the
interval crosses one and the stability conclusion is unresolved, even if
held-out posterior error is small.  If it is $.80$ with the same radius,
subunit amplification is certified on the tested prompt and evidence domain.
The minimax proposition says that no alternative method can uniformly eliminate
this grey region faster than the derivative-recovery boundary without
stronger structural assumptions.

\paragraph{Answer to Question 2.}
The inverse can be learned uniformly on its supported domain under the stated
smoothness and sampling conditions. Stability requires the harder task of
covering its derivative uniformly. A valid procedure must therefore be
allowed to return ``unresolved'' near the minimax boundary; low held-out error
alone is not a stability certificate.

\subsection{Partial identification under truncated probabilities}

\paragraph{Question 3: what remains identifiable when probabilities are
missing?}
If every semantic probability is observed, Questions 1 and 2 can lead to a
point estimate. If some probabilities are absent, a unique answer may no
longer be justified. This section replaces the unidentified point by the set
of posterior values compatible with the returned probabilities and the
recorded missing mass.

The point-recovery results above assume that the relevant semantic
probability vector is available.  Actual probability records may contain only
selected alternatives.  Treating every absent candidate as probability zero
would manufacture certainty.  The statistically honest response is to retain
all full probability vectors compatible with the returned coordinates and
missing mass, then map that set back to the posterior-coordinate space.

A robust procedure must handle settings in which some phrase probabilities
are unavailable without silently replacing them by zero. For an observed
record $\bm s$, define
$\mathcal J(\bm s)\subseteq\Delta_J$ as the set of full semantic laws compatible
with that record, including its reported missing mass. The
compatible posterior-coordinate set is
\[
 \mathcal C(\bm s)=
 \{\bm\xi\in D:\bm g_u(\bm\xi)\in\mathcal J(\bm s)\}.
\]
It is empty under model contradiction, a singleton under point
identification, and otherwise non-singleton.  In particular, an absent
candidate is not assigned probability zero.

\paragraph{Worked missing-mass example.}
Suppose the language service returns probabilities only for phrases mapped to
the first two of four semantic states.  Their grouped masses are $.55$ and
$.25$, leaving unreported mass $.20$.  The full state composition is
not $(.55,.25,0,0)$.  It belongs to
\[
 \{(.55,.25,a,.20-a):0\le a\le .20\}.
\]
The $\ell_1$ diameter of this compatible set is $.40$: one endpoint puts all
missing mass in state three and the other puts it in state four.  A midpoint
allocation $(.55,.25,.10,.10)$ is a convenient summary, but it is an
assumption rather than an identified probability vector.  The set
$\mathcal C(\bm s)$ transfers this ambiguity through the inverse map.

To state sharpness explicitly, let $R(\bm s)\subseteq\{1,\ldots,J\}$ index the
reported coordinates, let $q_j(\bm s)$ be their observed masses, and let
\[
 m(\bm s)=1-\sum_{j\in R(\bm s)}q_j(\bm s)
\]
be the reported unallocated mass. The compatible observation set is
\begin{equation}\label{eq:compatible-observation}
 \mathcal J(\bm s)=\left\{\bm p\in\Delta_J:
 p_j=q_j(\bm s)\ (j\in R(\bm s)),\quad
 \sum_{j\notin R(\bm s)}p_j=m(\bm s)\right\}.
\end{equation}
This set uses everything supplied by the truncated record and makes no
allocation among unreported states.

The identified-set logic is classical \citep{manski2003partial,
chernozhukov2007estimation,tamer2010partial}. The result below specializes it
to a truncated probability simplex and adds the missing-mass--inverse-modulus
diameter bound needed for the present observation channel.

\begin{theorem}[Sharp compatible posterior set]\label{thm:set}
Suppose the observation record $\bm s$ reports exactly the coordinates in
$R(\bm s)$ and the unallocated mass $m(\bm s)$, and suppose the observation operator
$\bm g_u:D\to\Delta_J$ is known. Then
\[
 \mathcal C(\bm s)=\{\bm\xi\in D:\bm g_u(\bm\xi)\in\mathcal J(\bm s)\}
\]
is the sharp identified set: every compatible posterior coordinate belongs to
$\mathcal C(\bm s)$, and every element of $\mathcal C(\bm s)$ generates the observed
record. If $\bm g_u$ is injective on $D$ with inverse $G_u$ satisfying
\[
 \|G_u(\bm p)-G_u(\bm p')\|
 \le C_u\|\bm p-\bm p'\|_1,
\]
then
\[
 \operatorname{diam}\{\mathcal C(\bm s)\}
 \le C_u\operatorname{diam}\{\mathcal J(\bm s)\cap\bm g_u(D)\}
 \le 2C_um(\bm s).
\]
When only one coordinate is unreported, its mass is determined by the simplex
constraint and the compatible observation set is a singleton.
\end{theorem}

The theorem gives the appropriate output when point recovery is unavailable.
Missing mass widens the observation set, while inverse conditioning determines
how strongly that width is amplified on the posterior scale. In the worked
example, the observation-set diameter is $.40$; if $C_u=1.5$, the posterior
diameter is at most $.60$. A reported point inside that set may be useful as a
summary, but it is not identified without an additional allocation rule.

The preceding results treat operator error, sampling error and coarsening in
separate steps. For deployment, however, the required object is one confidence
region that remains valid when all three occur together. The next theorem is
the paper's principal new inferential result. Unlike the classical ingredients
listed in Table~\ref{tab:theory-lineage}, it constructs and controls that
region for an estimated, set-valued probability experiment.

Let $T:D\to\mathcal M$ be the population observation operator into a metric
space $(\mathcal M,d_{\mathcal E})$, and let $\widehat T$ be its estimate from
calibration data. A new record $S$ determines a nonempty closed set
$\mathcal A(S)\subseteq\mathcal M$ of observation laws compatible with what
was returned. Under full observation, $\mathcal A(S)$ is a singleton; under
truncation it contains every completion allowed by the returned probabilities
and missing mass. Write
\[
 \operatorname{diam}_{\mathcal E}\{\mathcal A(S)\}
 =\sup_{P,P'\in\mathcal A(S)}d_{\mathcal E}(P,P').
\]
The term \emph{honest} follows the confidence-set literature: it refers to
coverage over the prespecified model class, not to an informal assertion of
trustworthiness \citep{gine2010confidence}.

\begin{theorem}[Principal result: unified honest recovery]
\label{thm:honest}
Let $\bm\xi_0\in D$ be the target. Suppose that, for known radii $a_n,b_R\ge0$,
\begin{align*}
 \Pr\!\left\{\sup_{\bm\xi\in D}
 d_{\mathcal E}\bigl(\widehat T(\bm\xi),T(\bm\xi)\bigr)
 \le a_n\right\}&\ge1-\alpha_T,\\
 \Pr\!\left\{\inf_{P\in\mathcal A(S)}
 d_{\mathcal E}\bigl(T(\bm\xi_0),P\bigr)\le b_R\right\}
 &\ge1-\alpha_S.
\end{align*}
No independence between these events is required. Define
\[
 \widehat{\mathcal C}_{\alpha}(S)
 =\left\{\bm\xi\in D:
 \inf_{P\in\mathcal A(S)}
 d_{\mathcal E}\bigl(\widehat T(\bm\xi),P\bigr)
 \le a_n+b_R\right\}.
\]
Then
\[
 \Pr\{\bm\xi_0\in\widehat{\mathcal C}_{\alpha}(S)\}
 \ge1-\alpha_T-\alpha_S.
\]
Moreover, on the intersection of the two displayed events, if
$\operatorname{diam}_{\mathcal E}\{\mathcal A(S)\}\le\delta(S)$, then
\[
 \operatorname{diam}\{\widehat{\mathcal C}_{\alpha}(S)\}
 \le
 \omega_D\{4a_n+2b_R+\delta(S)\}.
\]
If $\omega_D(t)\le C_Gt$ locally, the right-hand side is at most
$C_G\{4a_n+2b_R+\delta(S)\}$. With $a_n=b_R=0$, the region reduces to the
sharp inverse image $T^{-1}\{\mathcal A(S)\}$.
\end{theorem}

The theorem separates the four reasons a posterior region can remain wide:
the forward operator is estimated imprecisely ($a_n$), the new probability
record is noisy ($b_R$), the record admits several completions
($\delta(S)$), or the inverse is weakly conditioned ($\omega_D$). It therefore
provides one honest output rather than selecting post hoc between a point
interval and a compatible set. For top-$k$ probability records under
$\ell_1$ distance, Theorem~\ref{thm:set} gives
$\delta(S)\le2m(S)$, making the contribution of unreturned mass explicit.

This result deliberately stops at static identification. Propagating such
sets through a state transition, combining measurement error across time, and
translating posterior width into decision regret require a transition model
and loss function. Those are sequential-decision questions rather than
properties of the observation experiment alone.

\paragraph{Answer to Question 3.}
Missing probabilities generally identify a set, not a point. Its width is
controlled by two transparent quantities: how much probability mass is
unreported and how strongly the inverse amplifies differences in the full
semantic vector. Assigning zero to absent alternatives would hide both.

\section{Experimental design}

This section demonstrates the new recovery methodology and examines whether
its finite-sample behaviour is consistent with the guarantees derived above.
It has three goals. First, we check whether the implemented estimators exhibit
the rates, error amplification and coverage behaviour predicted by the theory.
Second, we examine whether the proposed diagnostics
distinguish settings in which posterior recovery is reliable from those in
which the observation channel is weakly identified, the returned probability
record is incomplete, or the learned inverse is locally unstable. Third, we ask whether the complete workflow
remains informative when the observation law is supplied by fitted language
models rather than by a known simulator. The experiments are therefore
demonstrations under increasingly realistic conditions, not empirical proofs
of mathematical results or a model leaderboard.

\begin{figure}[H]
\centering
\resizebox{\linewidth}{!}{
\begin{tikzpicture}[
  stage/.style={rounded corners=2pt,draw=blue!55!black,thick,
    fill=blue!7,text width=30mm,minimum height=24mm,
    align=center,font=\small,inner sep=4pt},
  compact/.style={rounded corners=2pt,draw=blue!55!black,thick,
    fill=blue!7,text width=25mm,minimum height=24mm,
    align=center,font=\small,inner sep=4pt},
  input/.style={rounded corners=2pt,draw=black!60,fill=white,
    text width=35mm,minimum height=11mm,align=center,font=\scriptsize,
    inner sep=3pt},
  reference/.style={rounded corners=2pt,draw=black!70,thick,fill=white,
    text width=35mm,minimum height=16mm,align=center,font=\small,
    inner sep=4pt},
  question/.style={rounded corners=2pt,draw=black!55,fill=white,
    text width=37mm,minimum height=14mm,align=center,font=\scriptsize,
    inner sep=3pt},
  arrow/.style={-{Latex[length=2.4mm]},thick,draw=black!75},
  compare/.style={<->,>=Latex,thick,dashed,draw=blue!60!black}
]
\path[use as bounding box] (-6.55,3.45) rectangle (6.55,-5.20);

\node[input] (design) at (-4.90,2.85)
 {Prompt $C_u(y)$ with evidence $y$};

\node[stage] (law) at (-4.90,.55)
 {\textbf{Conditional law over phrases}\\[1.5mm]
  {\tiny ``urgent review'': $.38$\\
  ``immediate escalation'': $.17$\\
  ``routine follow-up'': $.35$}};

\node[compact] (semantic) at (0,.55)
 {\textbf{Prompt-dependent semantic observation}\\[1.5mm]
  {\scriptsize urgent: $.55$\\
  routine: $.35$\\
  insufficient: $.10$}};

\node[compact] (estimate) at (4.90,.55)
 {\textbf{Estimated state posterior}\\[1.5mm]
  {\scriptsize urgent: $.60$\\
  routine: $.30$\\
  insufficient: $.10$}};

\node[reference] (target) at (4.90,-2.20)
 {\textbf{Independently defined reference posterior}\\
  $\bm\pi^\star(y)$};

\draw[arrow] (design) -- node[right=1mm,font=\scriptsize]
 {fitted language system} (law);
\draw[arrow] (law) --
 node[above=1.5mm,align=center,font=\scriptsize] {group by}
 node[below=1.5mm,align=center,font=\scriptsize] {declared\\meaning}
 (semantic);
\draw[arrow] (semantic) --
 node[above=1.5mm,align=center,font=\scriptsize] {held-out inverse}
 node[below=1.5mm,align=center,font=\scriptsize] {calibration}
 (estimate);
\draw[compare] (estimate) --
 node[left=1.2mm,font=\scriptsize] {held-out}
 node[right=1.2mm,font=\scriptsize] {assessment}
 (target);

\node[font=\small,text=blue!60!black] at (0,-3.15)
 {The statistical questions};

\node[question] (identify) at (-4.35,-4.25)
 {\textbf{Identification}\\
  Do distinct target posteriors induce distinct observable language laws?};
\node[question] (stable) at (0,-4.25)
 {\textbf{Learning and stability}\\
  Can the inverse be estimated uniformly without excessive amplification?};
\node[question] (partial) at (4.35,-4.25)
 {\textbf{Partial identification}\\
  What remains identified when only selected probabilities are returned?};

\draw[black!30] (law.south) |- (identify.north);
\draw[black!30] (semantic.south) -- (stable.north);
\draw[black!30] (target.south) |- (partial.north);
\end{tikzpicture}}
\caption{Operational sequence used in the language-model experiments. A
prompt containing evidence produces a conditional phrase law; declared
semantic grouping produces a prompt-dependent observation; and held-out
inverse calibration estimates the independently defined state posterior. The
three lower boxes identify the empirical questions tested by the design.
Numerical entries use the patient-diagnosis illustration from Section~1.}
\label{fig:workflow}
\end{figure}
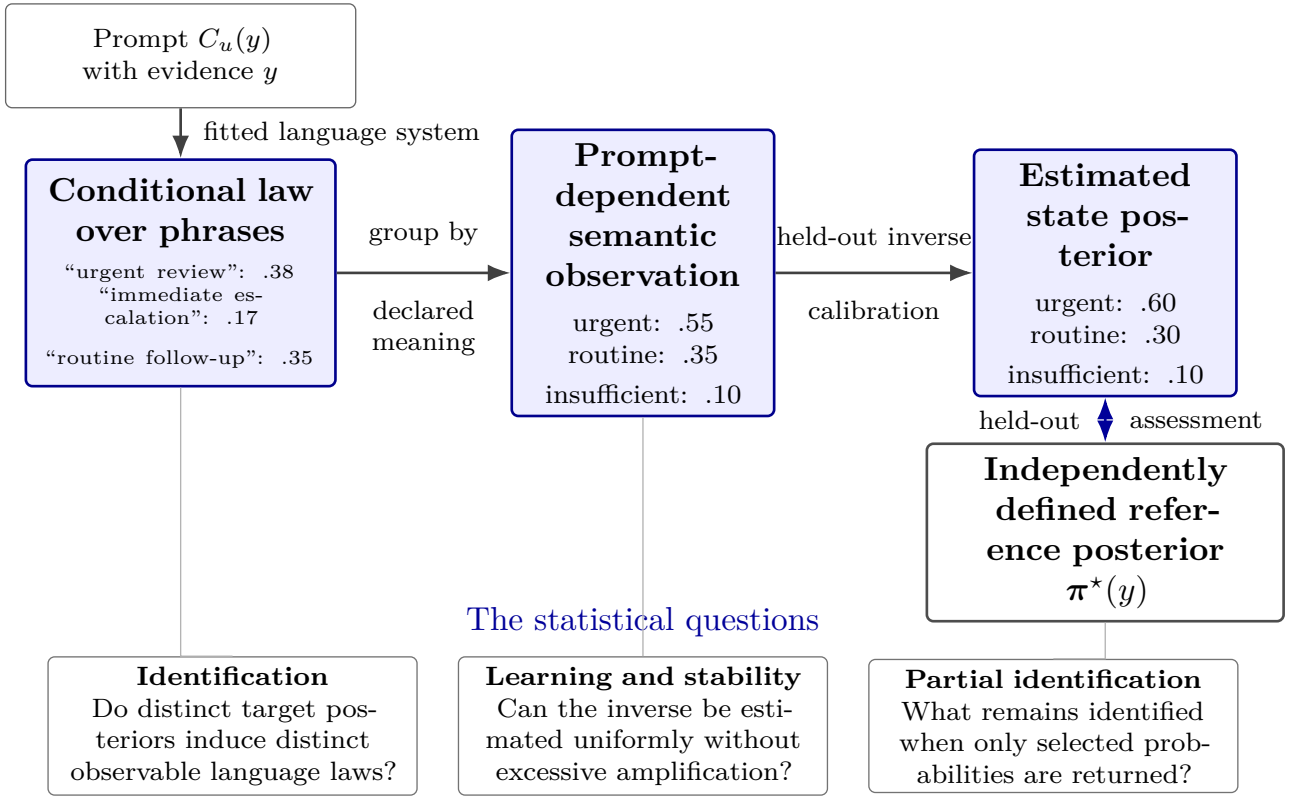

Figure~\ref{fig:workflow} translates the probability architecture in
Figure~\ref{fig:spaces} into the procedure implemented in the experiments.
Figure~\ref{fig:spaces} defines the inverse problem; Figure~\ref{fig:workflow}
shows how the experiments construct its observation and test identification,
stability and partial identification.

\subsection{From theorem to experimental prediction}

The simulations are designed around implications that could fail in finite
samples. For Question~1, Corollary~\ref{cor:replication} predicts an
$R^{-1/2}$ decline in recovery error when the inverse is identified and well
conditioned. The experiment therefore varies $R$ while holding the forward
map fixed, then separately weakens its smallest singular value. Support for
the theory requires both the predicted slope and greater error amplification
under the weaker inverse; an accurate fit at only one value of $R$ would not
be sufficient.

For Question~2, Theorem~\ref{thm:band} predicts a three-way outcome: certify
stability, certify excessive sensitivity, or remain unresolved. The design
places known derivative bounds below, near and above the threshold. A useful
procedure should increasingly classify cases away from the threshold while
retaining coverage and abstaining near it. The minimax experiment then gives
an oracle test full knowledge of the competing data-generating laws. Failure
below the predicted boundary therefore supports an intrinsic resolution limit
rather than an avoidable defect of the estimator.

For Question~3, Theorem~\ref{thm:set} predicts coverage of the generating
posterior and a compatible-set diameter that expands with omitted mass. The
experiment hides known coordinates through exact top-two reporting and checks
both properties. Coverage alone would be uninformative if obtained from an
arbitrarily enormous set; diameter is therefore reported alongside it.

The unified experiment then exercises the construction in
Theorem~\ref{thm:honest} with all principal
uncertainties present simultaneously. A scalar probability-valued forward map
is estimated within a certified uniform radius, a new probability is measured
from finitely many binomial repetitions, and its returned record is coarsened
to a known interval. The design varies the operator radius, repetition count,
coarsening diameter and forward slope. The theorem predicts coverage at or
above the declared level and a region that widens with all three error sources
and with weaker inverse conditioning.

The theorems require no empirical validation: their conclusions follow from
their assumptions. The experiments instead demonstrate that the new
methodology can be implemented and that its observed finite-sample behaviour
is consistent with the theory's distinct implications. A materially wrong
rate, systematic undercoverage, false declarations near the stability
boundary or failure of compatible-set coverage would expose a problem in the
implementation, approximation or claimed empirical scope.

\subsection{Applying the theory in practice}

Application requires the statistical experiment to be fixed before its final
evaluation. In practical order, the analyst declares the state space and
reference posterior; prespecifies phrase groups, prompt rules and the residual
category; records returned and missing probability mass; fits the inverse on
calibration cases; and evaluates conditioning, stability and coverage on
separate data. A point estimate is reported only when the observation is
sufficiently complete and the inverse passes its identification and stability
checks. Otherwise the appropriate output is a compatible set, an unresolved
stability decision, or abstention.

This is also the link to runtime governance. The theory does not attach a
generic trust label to a language model; it determines whether the particular
measurement is trustworthy for the declared inference and supplies an
observable trigger for continued use, review, recalibration or abstention.

Several design constraints follow. Calibration and deployment must share the
declared target, evidence class, prompt family, semantic partition and fitted
system; otherwise the forward law has changed. The calibration sample must
span the posterior directions that will be recovered, since held-out accuracy
cannot reveal a nearly invisible direction outside the sampled support.
Returned alternatives must retain their total mass and missingness status,
and repeated measurements must preserve scenario clusters. Finally, the
reference posterior must be constructed independently of the language
measurement, or apparent recovery becomes circular. These are restrictions on
what the resulting certificate means, not optional implementation details.

\subsection{Simulation and archived-study protocol}

The practical use case is motivated by financial-market monitoring.  An
analyst may receive evidence about earnings, economic activity, funding
conditions, credit spreads, and market volatility, yet require a probability
distribution over a smaller set of decision-relevant market regimes.  The
financial benchmark used below declares three such states: a favorable
regime, a weakening or mixed regime, and a severe-stress regime.  This setting
is useful because the same evidence can be expressed through many phrases,
while downstream statistical analysis requires comparable
state probabilities.  It provides a concrete application of the semantic
observation kernel without treating a language model's output as external
market truth.

\paragraph{A representative benchmark record.}
To make the controlled language data concrete, scenario 20260810-000 used the
prior $(.50,.30,.20)$ for Bull, Bear and Crisis and supplied the evidence
\begin{quote}\small
``Corporate earnings were strong; credit spreads remained tight; funding
conditions were easy; activity indicated expansion; market volatility was
elevated.''
\end{quote}
The system message was ``Return only the required structured candidate code.
Do not use outside information.'' The user message supplied the prior,
evidence, and the following frozen state definitions:
\begin{quote}\small
``A = favorable: strong or improving earnings and activity, easy funding,
tight credit and generally low volatility; B = weakening or mixed: softer
earnings and activity, normal-to-tighter funding, stable-to-wider credit and
elevated volatility; C = severe stress: weak earnings, contraction, stressed
funding, wide credit and extreme volatility. Select exactly one candidate
code using only these definitions, the prior and the evidence.''
\end{quote}
For one GPT-4.1-mini repetition, the raw candidate vector was
$(.3486233,.6513141,.0000626)$, with observed mass $1.00000003$ up to floating-
point error; the independently generated reference posterior was
$(.9927292,.0072636,.0000072)$. This deliberately large raw discrepancy
illustrates why held-out inverse calibration is required rather than equating
candidate probabilities with the target posterior.

\paragraph{Reproducible observation schema.}
Each archived row retains the scenario and presentation identifiers, repeat
number, requested and returned model, model fingerprint, provider response
identifier and timestamp, generated code, each candidate's raw log probability
and probability, the conditionally normalized vector, observed candidate mass,
parser method, parser cross-check and fallback status, and the reference
posterior. The complete frozen prompt, state definitions, prior, evidence,
tokenization of the candidate codes and partition assignment are stored in the
design files. Both model studies use the same 520 scenarios, five repetitions
per scenario and the fixed split 240 calibration-training, 80 validation, 100
conformal-calibration and 100 untouched test scenarios. The requested models
were GPT-4.1-mini and GPT-4o-mini-2024-07-18, with temperature zero, one
completion token and up to 20 returned token alternatives. Seeds
20260810--20260813, raw probability records, design and partition files,
analysis code and machine-readable outputs are included in the reproducibility
supplement. Replaying provider calls may not reproduce identical probabilities
after a service update; the archived raw records, rather than a later API
response, are therefore the authoritative inputs for reproducing the reported
tables. Appendix~\ref{app:example-data} displays one archived record in a
compact human-readable form.

The numerical study uses complementary sources of evidence.  Controlled
simulations provide known data-generating processes, making identification,
missing mass, derivative stability, and minimax separation directly
observable; Monte Carlo intervals quantify simulation uncertainty.  The
frozen financial study then uses archived candidate probabilities from two
fitted language models to test held-out posterior recovery, coverage, and
prompt-equivalent stability on a 520-scenario market-regime benchmark.  The
simulations test whether the mathematical conclusions are correct under their
stated assumptions, whereas the financial study tests whether the resulting
measurement procedure is empirically usable on its declared support.

The archived study connects the controlled theory to practice without
pretending that its unknown forward law is observable. Paired reference cases
permit estimation of the inverse; sampled rank and conditioning probe whether
posterior directions are visible; untouched test cases measure recovery;
conformal regions assess predictive coverage; and frozen prompt variants test
whether presentation changes behave as declared. Collectively these checks
can support use on the sampled domain. They cannot establish global
injectivity or justify transport to a new model, prompt family or evidence
process.

\paragraph{Which assumptions are tested?}
The simulations impose the semantic cells, reference posterior, support,
smoothness and observation mechanism, so these quantities are known rather
than inferred. They use independent calibration scenarios, matching
Assumption~\ref{ass:design}. The archived language study is likewise a
cross-section rather than a temporal process. A geometrically beta-mixing
extension is retained in the appendix as secondary theory, but no empirical
claim in this paper depends on it.

\begin{table}[t]\centering\footnotesize
\caption{Experimental design and its connection to the theory.  Each row
corresponds to a distinct inferential question; the fitted-language-model
study is held out from the simulations and uses no newly generated
probability measurements.}
\label{tab:study-design}
\begin{tabularx}{\linewidth}{p{.20\linewidth}p{.25\linewidth}p{.27\linewidth}X}
\toprule
Study component & Theoretical question & Statistical design & Primary
diagnostic\\
\midrule
Inverse recovery &
Does an identified inverse attain the predicted replication rate? &
Known scalar and two-dimensional injective maps; 400 Monte Carlo
repetitions across increasing $R$. &
Log--log RMSE slope and Monte Carlo interval.\\
Partial observation &
What remains identified when candidate mass is unobserved? &
Four-cell compositions with exact top-two reporting and compatible-set
propagation. &
Set coverage and $\ell_1$ diameter.\\
Stability certification &
Can subthreshold and excessive sensitivity be distinguished uniformly? &
Cubic Gaussian calibration surfaces with
$\kappa\in\{.75,.95,1.05,1.25\}$. &
Simultaneous coverage, correct decision, and abstention rates.\\
Minimax separation &
Is the derivative-resolution boundary intrinsic? &
Oracle likelihood-ratio tests under alternatives below, at, and above the
$n^{-1/5}$ boundary. &
Classification power by sample size and separation regime.\\
Financial application &
Do observable language probabilities recover the declared posterior on
held-out cases? &
Two frozen 520-scenario market-regime studies, five repeated measurements,
and disjoint calibration, validation, conformal, and test partitions. &
Test Jensen--Shannon divergence, 90\% coverage, conditioning, and
prompt-equivalent stability.\\
\bottomrule
\end{tabularx}
\end{table}

Every simulation corresponds to a preceding theorem and uses the fixed seed
20260724.  No fitted-language-system calls are made.  First, scalar and
two-dimensional posterior coordinates are mapped injectively to,
respectively, three- and four-cell semantic laws and estimated from repeated
multinomial observations.  Second, four-cell compositions are subjected to
exact top-two reporting; the unreturned mass is allocated over all compatible
omitted cells. Third, correctly specified cubic Gaussian conditional-mean
calibration surfaces with $\kappa\in\{0.75,0.95,1.05,1.25\}$ are fitted and assessed using
the exact simultaneous derivative band in
Corollary~\ref{cor:gaussianband}.  Finally, smooth localized alternatives
are placed below, at, and above the $n^{-1/5}$ boundary corresponding to
$s=2,d=1$.  An oracle likelihood-ratio test is used so that failure below the
boundary cannot be attributed to an inefficient classifier.

A targeted ablation then holds the two-dimensional estimator fixed while
altering one feature of the data-generating process at a time: replication,
minimum singular value, semantic-cell contamination, and a prompt-specific
logit offset.  A final row removes that offset using a correction frozen
before evaluation.  This design distinguishes sampling variability, weak
identification, semantic misspecification, and presentation bias.  All
reported proportion intervals are Wilson intervals; RMSE intervals apply a
delta-method calculation on the Monte Carlo mean-square scale.

\section{Results and discussion}

Section~5 contains two distinct forms of evidence. The controlled-simulation
subsection makes no language-model calls: it demonstrates inverse recovery, the
stability boundary and partial identification when the data-generating law is
known. The archived-language-measurement subsection is the only empirical
component that uses fitted language models. There, raw candidate probabilities
previously obtained from GPT-4.1-mini and GPT-4o-mini are calibrated against
independently generated reference posteriors and evaluated on untouched test
scenarios. Thus the simulations show that the new methodology behaves
consistently with its guarantees under controlled designs, while the
fitted-model study asks whether the proposed inverse measurement is feasible
with actual language probabilities. Within those two parts, the results follow
the three theoretical questions in order and then assess the unified region
that combines inverse recovery with partial observation.

\subsection{Controlled simulations}

\begin{table}[t]\centering\small
\caption{Theory-guided demonstrations of the proposed methodology. Monte Carlo repetitions are
400, with 80 repetitions in each stability cell. Parenthetical intervals are
95\% Monte Carlo intervals and are distinct from the simultaneous
derivative confidence band being evaluated.}
\label{tab:simulation}
\begin{tabularx}{\linewidth}{p{.28\linewidth}p{.22\linewidth}X}
\toprule
Claim & Target & Result\\\midrule
Replicated inverse rate &
$R^{-1/2}$ &
Log--log RMSE slope $-0.485$; RMSE falls from $0.314$ at $R=25$ to
$0.042$ at $R=1600$ (Monte Carlo interval $0.039$--$0.045$).\\
Multidimensional inverse &
$R^{-1/2}$ in $\RR^2$ &
Four-cell recovery has slope $-0.520$; RMSE falls from $0.503$ at $R=50$
to $0.057$ at $R=3200$ ($0.053$--$0.061$).\\
Top-two partial identification &
Generating law is covered &
Coverage is $400/400$ in every concentration stratum (Wilson lower endpoint
$0.990$); mean set diameter equals twice the observed omitted mass.\\
Unified honest recovery &
$95\%$ coverage and diameter bound &
Coverage is $0.990$--$1.000$ across 81 operator-error, replication,
coarsening and conditioning cells; every mean diameter is below its theorem
bound.\\
Stability gate &
Interval coverage and safe abstention &
Simultaneous coverage is $0.988$--$1.000$ across cells; no false
classification occurred (cellwise Wilson upper endpoint $0.046$);
unresolved decisions concentrate near $\kappa=1$.\\
Minimax boundary &
$n^{-1/5}$ &
Below-boundary oracle accuracy remains far from one ($0.660$ at $n=100$ and
$0.615$ at $n=2000$); at-boundary accuracy also remains below one;
above-boundary accuracy is $1.00$ in these cells.\\
\bottomrule
\end{tabularx}
\end{table}

\begin{figure}[!htbp]\centering
\includegraphics[width=.72\linewidth]{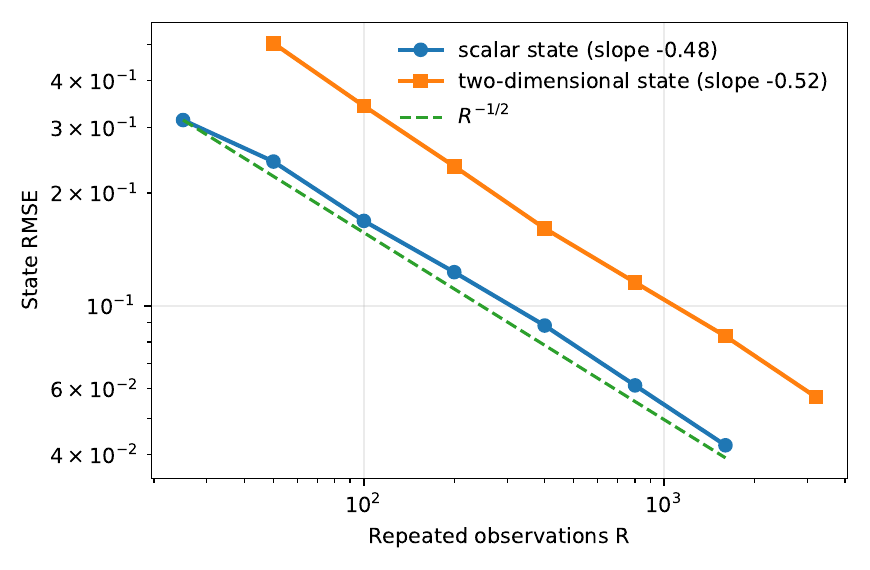}
\caption{\textbf{Question 1: inverse recovery.} This figure investigates the
$R^{-1/2}$ replication rate in Corollary~\ref{cor:replication} and shows that
both estimated slopes are close to $-1/2$. The higher
two-dimensional level reflects additional coordinates rather than a failure
of the predicted rate.}
\label{fig:inverse-rate}
\end{figure}

\begin{figure}[!htbp]\centering
\includegraphics[width=.94\linewidth]{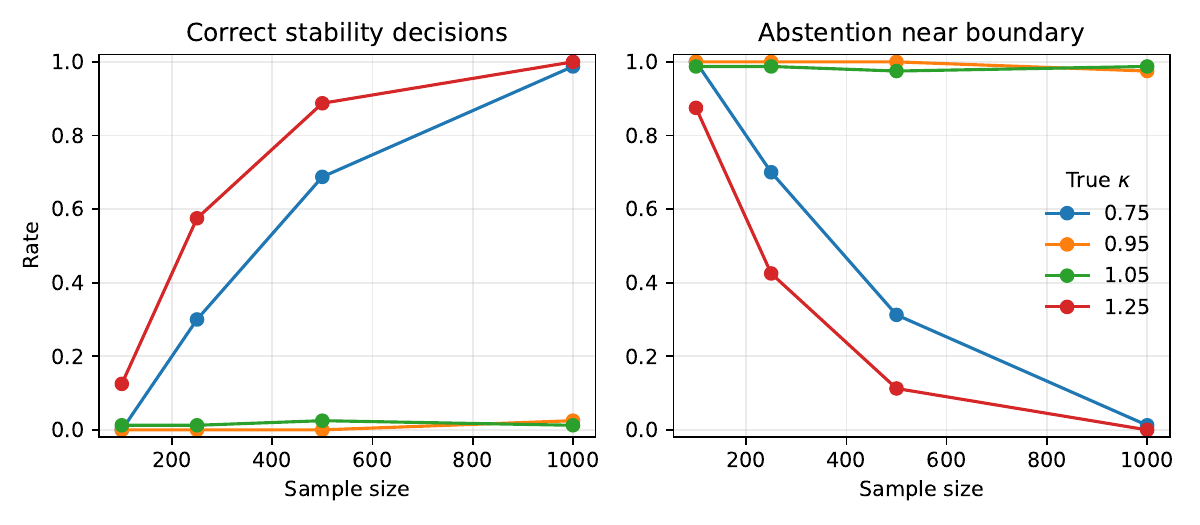}
\caption{\textbf{Question 2: stability certification.} This figure
investigates classification and abstention under the simultaneous derivative
band. It shows that correct decisions increase with sample size away from
$\kappa=1$, while
unresolved decisions near $.95$ and $1.05$ preserve the advertised coverage
rather than conceal errors.}
\label{fig:stability-gate}
\end{figure}

\begin{figure}[!htbp]\centering
\includegraphics[width=.72\linewidth]{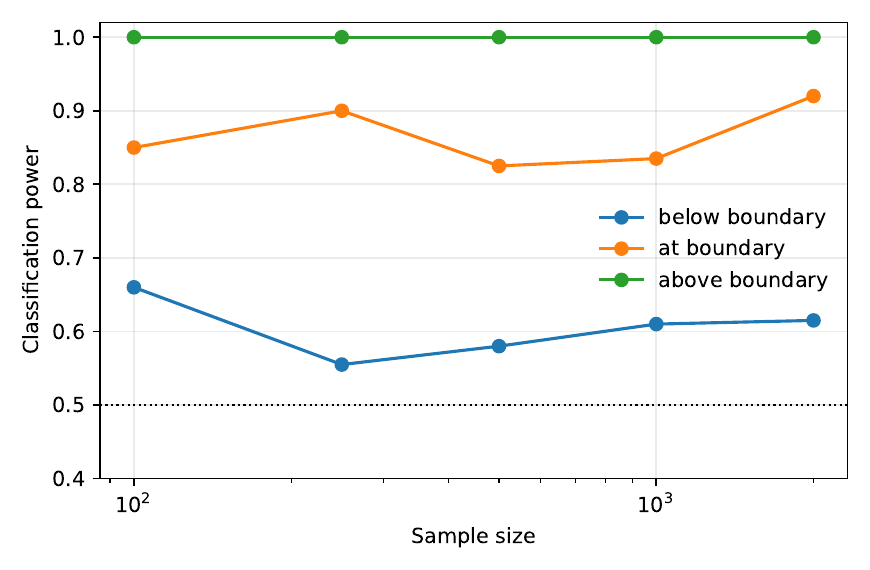}
\caption{\textbf{Question 2: minimax resolution.} This figure investigates
the $n^{-1/5}$ boundary in Proposition~\ref{thm:minimax} using an oracle test. It
shows that accuracy remains bounded away from one below the boundary, is not uniformly one
at the boundary, and reaches one above it in the reported cells.  This supports
an intrinsic resolution limit rather than an estimator-specific failure.}
\label{fig:minimax-boundary}
\end{figure}

\begin{figure}[!htbp]\centering
\includegraphics[width=.94\linewidth]{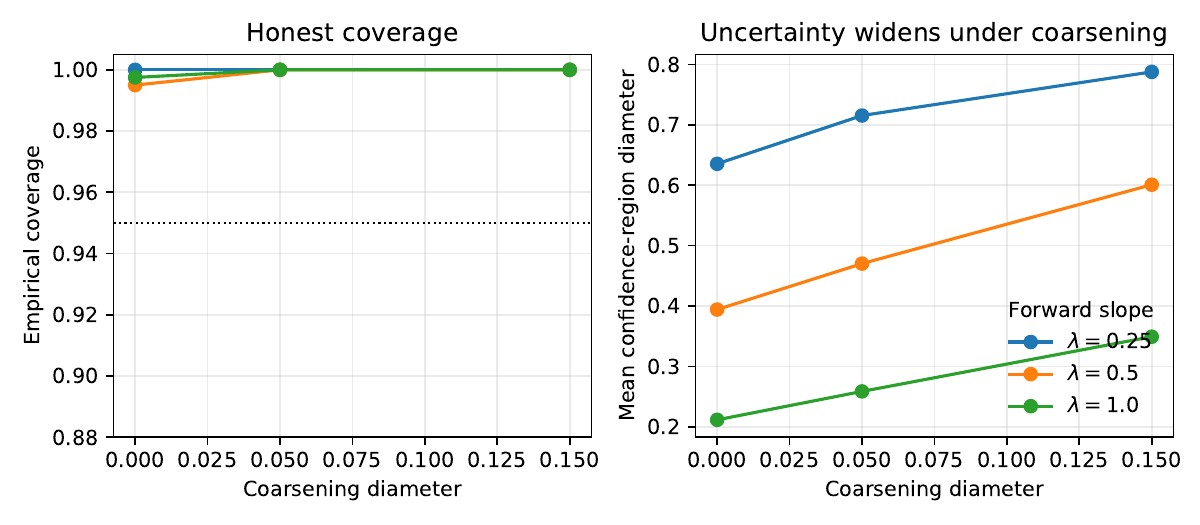}
\caption{\textbf{Unified honest recovery.} The left panel shows empirical
coverage for $a_n=.01$ and $R=200$; the dotted line is the nominal $.95$
level. The right panel shows that confidence regions widen as the compatible
observation record is coarsened and as the forward slope $\lambda$ decreases.
The complete table covers all 81 combinations of operator radius, replication,
coarsening and conditioning.}
\label{fig:honest-recovery}
\end{figure}

\paragraph{Reading the simulation table.}
No language model is used in Table~\ref{tab:simulation} or
Figures~\ref{fig:inverse-rate}, \ref{fig:stability-gate},
\ref{fig:minimax-boundary} and~\ref{fig:honest-recovery}. Their data are
generated from known probability laws so that the target, forward map,
conditioning and missing mass are observed exactly. The 400 Monte Carlo
replications give useful precision for RMSE and coverage comparisons; the 80
replications in each stability cell are less informative about very rare
errors, which is why Wilson intervals accompany the reported proportions.
The purpose is isolation: each experiment removes the ambiguity of an unknown
language system and asks whether one mathematical mechanism appears in finite
samples.

The scalar and two-dimensional slopes are close to $-1/2$, so doubling the
number of repeated observations reduces RMSE by roughly $1/\sqrt2$ when the
inverse is well conditioned.  Exact compatible-set coverage is achieved by
widening the set as missing mass grows, not by pretending truncation is
innocuous.  The derivative gate is deliberately conservative near
$\kappa=1$: it makes no false classification in these cells, but often
returns unresolved for $\kappa=.95$ or $1.05$.  The minimax experiment then
shows why this difficulty persists even for an oracle test.

\paragraph{Figure~\ref{fig:inverse-rate}: Question 1---can an identified
inverse recover the target at the predicted rate?}
Corollary~\ref{cor:replication} predicts that replication error decreases as
$R^{-1/2}$ when the inverse is locally well conditioned.  The figure tests
this prediction in one- and two-dimensional target spaces. The experiment
deliberately assumes a known injective forward map, exact semantic cells,
complete probability mass and conditionally independent multinomial
replicates. Those assumptions remove semantic error, prompt variation,
partial observation and forward-map estimation so that replication error is
the only changing quantity.

The observed slopes, $-0.485$ and $-0.520$, closely match the predicted
$-1/2$. The higher two-dimensional curve is expected because two coordinates
must be recovered; the slopes rather than the intercepts assess the method's
finite-sample scaling.
This is strong evidence that the implemented inverse exhibits the predicted
replication behavior under controlled conditions. It does not show that a
real language-derived forward map is injective, nor does it test dependent
scenarios or errors in semantic grouping. Those failures are examined
separately rather than being hidden inside this rate experiment.

\paragraph{Figure~\ref{fig:stability-gate}: Question 2---can instability be
detected without declaring borderline cases safe?}
Theorem~\ref{thm:band} and Corollary~\ref{cor:gaussianband} require a
simultaneous derivative band.  The figure tests whether the resulting gate
classifies surfaces away from the stability boundary while abstaining when
the data cannot separate $\kappa$ from one. Again, no language model is used.
The calibration surface is a correctly specified cubic function, the design
is fixed and full rank, and the errors are independent Gaussian. These strong
assumptions make the appendix's exact finite-sample band available and isolate
the decision rule from model misspecification.

The resulting pattern tells a coherent story. Away from one, correct
classification improves with sample size. Near $.95$ and $1.05$, the band
often crosses the threshold and the method reports ``unresolved'' rather than
forcing a favorable conclusion. Simultaneous coverage of $.988$--$1.000$ and
the absence of false classifications in the reported cells are consistent
with the finite-sample construction, although 80 replications per cell cannot
rule out a small error probability. The figure demonstrates the specialized
implementation used here; application to language data requires the more
general simultaneous-band conditions in Theorem~\ref{thm:band}, not an
assumption that an unknown inverse is exactly cubic and Gaussian.

\paragraph{Figure~\ref{fig:minimax-boundary}: Question 2 continued---is the
difficulty near the stability boundary intrinsic?}
Proposition~\ref{thm:minimax} implies that no method can uniformly resolve
derivative alternatives below the $n^{-1/5}$ boundary in this design.  The
oracle experiment removes estimator inefficiency as an explanation: if even
the oracle cannot separate the alternatives, the limitation belongs to the
statistical problem. The data are synthetic localized alternatives with known
likelihoods; no fitted language model or estimated inverse enters the test.

Accuracy remains bounded away from one below and at the boundary, and reaches
one above it in the reported cells. This is the
finite-sample pattern predicted by the lower bound and explains the
abstentions in Figure~\ref{fig:stability-gate}: some borderline cases are
statistically unresolved, not merely estimated badly. The experiment is not
an empirical proof of a minimax theorem, and its finite grid does not establish
the sharp constant or behavior in every dimension. Its value is that the
qualitative boundary remains visible even after giving the test oracle
information unavailable in practice.

\paragraph{Question 3---what remains identifiable after truncation?}
This question has a set-valued rather than graphical answer.  The top-two row
of Table~\ref{tab:simulation} demonstrates the construction in
Theorem~\ref{thm:set}: the compatible set
contains the generating posterior in all 400 replications, while its diameter
increases with omitted mass.  Thus coverage is obtained by retaining
unresolved probability mass, not by converting missing values into zeros.
Here too the probability vectors and truncation rule are simulated rather
than produced by an LLM. Exact knowledge of the omitted mass isolates the
geometry of partial identification. In practice, a deployed language service
must return the omitted mass, or enough probability information to bound it
defensibly; otherwise the compatible set cannot inherit the same guarantee.
This requirement is realistic for a constrained finite response grammar when
the service exposes the conditional token probabilities needed to score every
admissible phrase, or explicitly reports residual probability mass. It is not
automatically satisfied by a top-$k$ list from unrestricted free-text output,
for which the unreturned phrase mass may remain unidentified.

This limitation does not invalidate the archived language-model study. That
study used a constrained finite response grammar, requested raw token
probabilities for more alternatives than the declared grammar contained, and
retained those probabilities before normalization. Every declared candidate
was returned, and the minimum observed candidate mass was approximately
$0.9999998$. The experiment therefore provides near-complete observation of
its task-specific candidate law, not of the model's unrestricted free-text
distribution. Its recovery and coverage claims remain conditional on the
frozen grammar, prompt family, fitted model and evidence domain. By contrast,
the partial-identification simulation deliberately withholds some component
probabilities while retaining the total mass assigned to the omitted
components. It then constructs the complete set of posterior values consistent
with the reported probabilities and that omitted mass, rather than assigning
zero or an arbitrary allocation to the unreported alternatives.

\paragraph{Figure~\ref{fig:honest-recovery}: do the separate uncertainties
combine into an honest recovery region?}
This experiment directly demonstrates the recovery construction in
Theorem~\ref{thm:honest}. The scalar
forward law is $g_\lambda(\xi)=0.5+\lambda\xi$ on
$D=[-0.4,0.4]$, with $\lambda\in\{0.25,0.5,1\}$. A random fitted operator is
perturbed within a known uniform radius $a_n\in\{0,.01,.03\}$. A new
probability is estimated from $R\in\{50,200,800\}$ binomial repetitions, with
$b_R$ set by the $95\%$ Hoeffding radius, and the returned record is an
interval of diameter $\delta\in\{0,.05,.15\}$. These choices form 81 cells,
each evaluated in 400 Monte Carlo replications.

Empirical coverage ranges from $.990$ to $1.000$, so no cell falls below the
nominal $.95$ level at the reported Monte Carlo resolution. Every mean region
diameter is below the theorem's bound. The right panel shows the mechanism:
for $a_n=.01$ and $R=200$, increasing coarsening diameter from $0$ to $.15$
increases mean region diameter from $.394$ to $.601$ when $\lambda=.5$.
Smaller $\lambda$ produces wider regions because its inverse Lipschitz
constant is larger. The conservatism is expected from Hoeffding's inequality,
the union bound and the worst-case operator radius; it is the price of honest
rather than pointwise or post-selected coverage.

\begin{table}[!htbp]\centering\footnotesize
\caption{Targeted robustness study for the two-dimensional, four-cell inverse
at 400 Monte Carlo repetitions.  RMSE intervals quantify Monte Carlo error,
not posterior uncertainty.  The inverse is held fixed except in the final row,
which applies a correction estimated and frozen before evaluation.}
\label{tab:robustness}
\begin{tabularx}{\linewidth}{p{.31\linewidth}rrrrX}
\toprule
Condition & $R$ & $\sigma_{\min}$ & RMSE (95\% MC) & Bias norm & Interpretation\\
\midrule
Baseline &
400 & 1.129 & .165 (.155, .175) & .006 &
Well-conditioned recovery.\\
Fewer repeated observations &
100 & 1.129 & .327 (.308, .346) & .010 &
Sampling error rises at approximately the square-root rate.\\
Weak identification &
400 & .290 & .503 (.467, .536) & .008 &
A small singular value amplifies probability error despite little bias.\\
5\% semantic-cell contamination &
400 & 1.129 & .186 (.175, .196) & .037 &
An incorrect grouping introduces persistent error and bias.\\
Unadjusted prompt offset &
400 & 1.129 & .235 (.222, .248) & .166 &
Presentation shift is confounded with state movement.\\
Frozen prompt-offset correction &
400 & 1.129 & .174 (.162, .186) & .008 &
Pre-test correction largely restores baseline recovery.\\
\bottomrule
\end{tabularx}
\end{table}

Table~\ref{tab:robustness} shows why one aggregate accuracy number is
insufficient.  Reducing the number of repeated observations mainly increases
variance; weakening the observation operator amplifies variance through the
inverse modulus; semantic contamination creates model misspecification; and
an unadjusted prompt offset creates systematic bias.  The final row is not a
post-hoc repair: the offset is treated as a calibration parameter fixed before
the test observations are evaluated.  Its recovery toward the baseline
supports the paper's separation of presentation effects from evidence-driven
posterior movement.

The top-two experiment deserves care in interpretation.  Its exact coverage
is not evidence that truncation is harmless: the average compatible-set
$\ell_1$ diameter increases from $0.141$ to $0.700$ as the Dirichlet
concentration makes the omitted cells less negligible.  Coverage is preserved
because uncertainty is honestly widened.  Replacing absent probabilities by
zero would instead produce narrower but invalid conclusions.

\subsection{Archived language measurements}

\paragraph{Where the fitted language models enter.}
This is the only empirical component that uses probabilities produced by
fitted language models. All preceding rate, stability, minimax and
partial-identification experiments are controlled simulations. Keeping those
roles separate matters: simulation makes the target and forward law known,
whereas this study asks whether the workflow remains informative when the
forward law is unknown and must be estimated.

Controlled simulations can demonstrate the methodology against theoretical
implications because the true observation operator is known. They cannot show
that actual language probabilities carry
recoverable state information.  The archived illustration serves that second,
more modest purpose.  Its question is not whether a language system knows the
market.  It is whether, on a frozen controlled benchmark with a known
reference posterior, observable candidate probabilities support held-out
recovery and uncertainty coverage.

The empirical illustration reuses two frozen controlled studies, each with
520 scenarios, five repeated measurements per scenario, and disjoint
calibration, validation, conformal and 100-scenario test partitions.  The
reference posteriors come from the prespecified synthetic evidence-generating
model; the fitted language systems supply observable candidate probabilities.
The illustration therefore evaluates recovery from actual language
probabilities while retaining a known posterior target.  It is not evidence
that the reference model describes an external market truth.

\paragraph{Data quality and scope.}
The principal strengths are frozen probability records, disjoint calibration,
validation, conformal and test partitions, and five repetitions per scenario.
These choices reduce post-selection bias and separate within-scenario
variation from held-out recovery. The limitations are equally material: each
model has only 100 final-test scenarios, only two fitted systems are studied,
the application is one market-regime benchmark, and the reference posterior
comes from a controlled synthetic evidence model. The repetitions do not form
a long temporal sequence and therefore provide no test of service drift or
the beta-mixing extension. The study is consequently an informative
feasibility check, not a broad claim about arbitrary LLMs or financial data.

\begin{table}[t]\centering\footnotesize
\caption{Held-out empirical illustration from frozen language measurements.
JS denotes Jensen--Shannon divergence. Means average recovery error over all
held-out measurements, not over state coordinates; confidence intervals are
clustered by the 100 test scenarios. Coverage is for a nominal 90\% conformal
region.}
\label{tab:empirical}
\begin{tabular}{lrrrrr}
\toprule
Fitted system & Test $n$ & Raw mean JS & Calibrated mean JS (95\% interval) & Radius & Coverage\\
\midrule
GPT-4.1-mini & 100 & .1898 & .0390 (.0295, .0498) & .1371 & .940\\
GPT-4o-mini & 100 & .1513 & .0366 (.0286, .0454) & .0922 & .900\\
\bottomrule
\end{tabular}
\end{table}

For clarity, Table~\ref{tab:empirical-samples} records the observational unit
behind the empirical summaries. Repetitions are repeated language
measurements of the same scenario and are not treated as independent
scenarios.

\begin{table}[t]
\caption{Samples used in the frozen language-model illustration.}
\label{tab:empirical-samples}
\centering
\small
\begin{tabular}{lrrl}
\toprule
Sample & Scenarios & Language measurements & Role\\
\midrule
In-sample & 240 & 1,200 & Fit the inverse map\\
Out-of-sample & 100 & 500 & Untouched evaluation\\
\bottomrule
\end{tabular}
\end{table}

Table~\ref{tab:state-composition} complements the recovery errors in
Table~\ref{tab:empirical} by showing how calibration changes the average
state composition. The averaging is over scenarios, never over the three
states: each column reports the sample mean of one state probability. The
intervals are scenario-bootstrap intervals and therefore preserve the five
repeated language measurements within each scenario. Agreement of these
averages is only a distribution-level diagnostic; it does not replace the
observation-level recovery and coverage results above.

\begin{table}[t]
\caption{Mean state probabilities with scenario-bootstrap 95\% intervals.
$A$ denotes Bull (favourable market regime), $B$ Bear (weakening or mixed
regime), and $C$ Crisis (severe-stress regime). In-sample results use the 240
calibration-training scenarios; out-of-sample results use the 100 untouched
test scenarios. Each scenario has five repeated language measurements. Means
and intervals are calculated over scenarios separately for each state, not
over states.}
\label{tab:state-composition}
\centering
\scriptsize
\setlength{\tabcolsep}{3.5pt}
\begin{tabular}{lrrr}
\toprule
Source & $P(A)$: mean [lower, upper] & $P(B)$: mean [lower, upper] & $P(C)$: mean [lower, upper]\\
\midrule
\multicolumn{4}{l}{\emph{In-sample: 240 calibration-training scenarios}}\\
GPT-4.1-mini, raw        & .1642 [.1208, .2125] & .4625 [.4007, .5203] & .3733 [.3146, .4301]\\
GPT-4.1-mini, calibrated & .3364 [.2917, .3847] & .4749 [.4405, .5102] & .1887 [.1571, .2194]\\
\cmidrule(lr){1-4}
GPT-4o-mini, raw         & .1491 [.1054, .1972] & .5900 [.5337, .6469] & .2608 [.2097, .3079]\\
GPT-4o-mini, calibrated  & .3446 [.3023, .3899] & .4718 [.4365, .5086] & .1836 [.1479, .2206]\\
\cmidrule(lr){1-4}
Reference posterior      & .3892 [.3466, .4319] & .4469 [.4102, .4843] & .1639 [.1292, .1978]\\
\midrule
\multicolumn{4}{l}{\emph{Out-of-sample: 100 untouched test scenarios}}\\
GPT-4.1-mini, raw        & .1842 [.1129, .2630] & .5178 [.4200, .6079] & .2980 [.2191, .3826]\\
GPT-4.1-mini, calibrated & .3505 [.2832, .4203] & .5000 [.4415, .5570] & .1495 [.1109, .1947]\\
\cmidrule(lr){1-4}
GPT-4o-mini, raw         & .1584 [.0954, .2337] & .6445 [.5540, .7322] & .1970 [.1343, .2751]\\
GPT-4o-mini, calibrated  & .3761 [.3094, .4403] & .4775 [.4213, .5347] & .1465 [.1020, .2038]\\
\cmidrule(lr){1-4}
Reference posterior      & .4001 [.3336, .4674] & .4459 [.3843, .5075] & .1540 [.1068, .2100]\\
\bottomrule
\end{tabular}
\end{table}

\paragraph{Table~\ref{tab:empirical}: held-out recovery and uncertainty.}
Across held-out measurements, calibration reduced mean JS divergence by
79.5\% for GPT-4.1-mini and 75.8\% for GPT-4o-mini. Modal-state accuracy rose
from 65.2\% to 80.8\% and from 72.0\% to 79.8\%, respectively. These are
averages across held-out observations; each divergence compares two complete
three-state probability vectors. Nominal 90\% regions covered 94 and 90 of
the 100 test-scenario reference posteriors. Thus the improvement is not merely
an in-sample fit statistic: it survives untouched evaluation and is reported
with predictive uncertainty. Both calibrations had full sampled design rank
and condition numbers near 20. Raw-response parser cross-checks passed and no
missing candidate was silently replaced by zero. These results support
recovery on the frozen experimental support, but not global injectivity or
transport to arbitrary prompts, models, domains or evidence processes.

\paragraph{Table~\ref{tab:empirical-samples}: what is independent.}
The table prevents the five repetitions from being mistaken for five new
scenarios. The inverse is fitted on 240 independent scenario clusters and
evaluated on 100 untouched clusters; the 1,200 and 500 language measurements
quantify within-scenario variation. Consequently, intervals and coverage are
clustered at scenario level. Counting all repeated measurements as independent
would understate uncertainty and make the apparent sample size misleading.

\paragraph{Table~\ref{tab:state-composition}: bias, variance and overfitting.}
Before calibration, both systems place too little average mass on Bull and too
much on Bear relative to the reference distribution. Calibration moves all
three state means toward their reference values in both samples. For example,
GPT-4.1-mini's out-of-sample Bull mean moves from $.1842$ to $.3505$ against a
reference mean of $.4001$; GPT-4o-mini moves from $.1584$ to $.3761$. The
calibrated in-sample and out-of-sample discrepancies are similar in direction
and magnitude, so the table shows no conspicuous train--test deterioration.
It does not prove absence of overfitting: averages can conceal offsetting
scenario-level errors, and the smaller 100-scenario test sample naturally has
wider intervals than the 240-scenario training sample. Table~\ref{tab:empirical}
therefore supplies the stronger check through observation-level held-out JS
error and conformal coverage.

This comparison also exposes the relevant bias--variance trade-off. A more
flexible inverse could remove additional calibration bias but would estimate
more degrees of freedom from only 240 training scenarios and could amplify
language-measurement noise, especially with condition numbers near 20. The
prespecified affine inverse accepts some residual bias in exchange for lower
estimation variance and transparent conditioning. Validation fixes that
choice before testing, while the conformal layer covers remaining recovery
error. The evidence supports that trade-off on this benchmark; it does not
establish that the same calibration complexity is optimal elsewhere.

\begin{figure}[!htbp]\centering
\includegraphics[width=.86\linewidth]{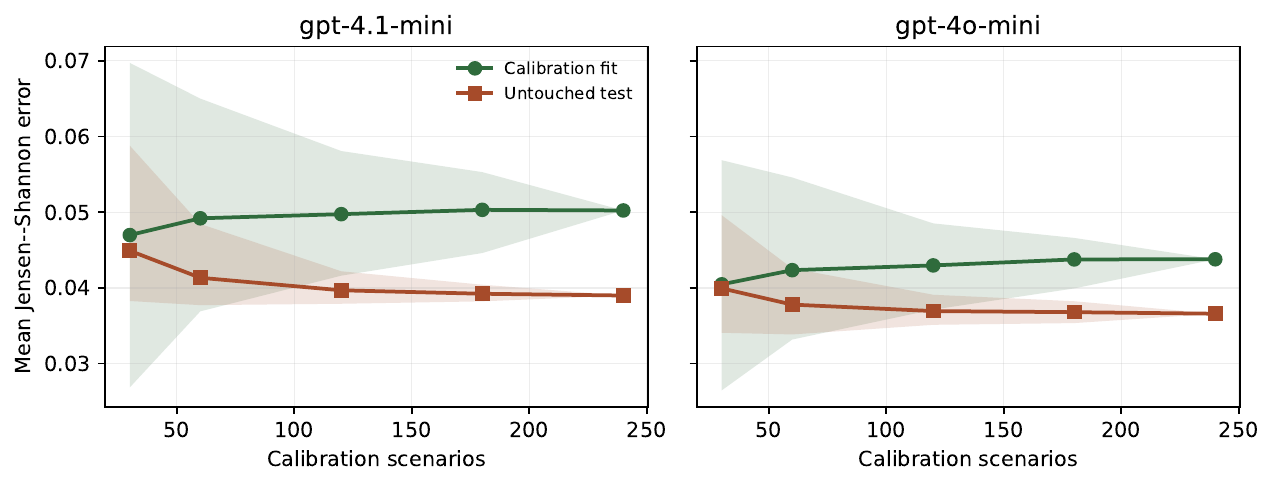}
\caption{\textbf{Empirical calibration learning curve.} Mean JS error for
affine inverses fitted to increasing numbers of calibration scenarios. Bands
show the 2.5th and 97.5th percentiles over 200 scenario-level subsamples; the
full-sample endpoint is unique. Untouched-test error decreases and then
flattens rather than increasing, indicating diminishing returns without an
observed overfitting reversal.}
\label{fig:empirical-learning}
\end{figure}

\paragraph{Figure~\ref{fig:empirical-learning}: does error decrease with more
calibration scenarios?}
Yes, but with diminishing returns. The same affine inverse is refitted on 200
scenario-level subsamples at $n=30,60,120$ and $180$; the $n=240$ point uses
the full calibration sample. All five repetitions from a scenario remain
together, and every fitted inverse is evaluated on the same untouched
100-scenario test partition. GPT-4.1-mini test JS falls from $.0449$ at 30
scenarios to $.0390$ at 240, while GPT-4o-mini falls from $.0399$ to $.0366$.
Most of the gain is achieved by about 120 scenarios. Test error does not turn
up as sample size grows, so the plot gives no evidence of classical overfitting
within the affine class. The flattening instead suggests that benchmark size
is only part of the remaining error: additional scenarios should reduce
estimation variance, but systematic approximation error and inverse
conditioning are increasingly important. The benchmark remains limited,
however. With 240 calibration scenarios and only 100 untouched test
scenarios, it cannot locate the asymptotic error floor or reliably distinguish
the small improvement that a substantially larger calibration sample might
produce. The result should therefore be read as a feasibility demonstration,
not evidence that recovery error has converged to its population limit.

\paragraph{What the empirical numbers mean.}
Coverage $.940$ on one 100-scenario test partition means that the nominal
90\% conformal region contained the reference posterior in 94 scenarios.
Coverage $.900$ means 90 of 100. These are finite test counts, not proof of
universal validity. Condition numbers near 20 warn that modest probability
errors can be amplified, which is why the paper reports coverage and parser
completeness alongside mean divergence.

\paragraph{Connection back to the running example.}
The empirical workflow has the same structure as triage.  Evidence and a
prespecified reference model generate a target posterior.  Candidate phrases
are assigned observable probabilities, grouped into semantic states, and
calibrated on one partition.  Only after every choice is frozen are posterior
recovery, prompt equivalence, and uncertainty coverage evaluated on held-out
scenarios.  This order prevents the semantic map or prompt family from being
chosen to flatter test outcomes.

\subsection{What the evidence establishes}

The experiments are most easily read as demonstrations of three capabilities
of the proposed methodology rather than as one aggregate performance claim.
Table~\ref{tab:simulation} and Figure~\ref{fig:inverse-rate} establish the
controlled recovery rates; Figures~\ref{fig:stability-gate} and
\ref{fig:minimax-boundary} separate certifiable stability from intrinsically
unresolved cases; Figure~\ref{fig:honest-recovery} and the partial-observation
row of Table~\ref{tab:simulation} assess honest set-valued recovery;
Table~\ref{tab:robustness} separates variance, weak identification and bias;
and Tables~\ref{tab:empirical}--\ref{tab:state-composition} ask whether the
same measurement logic remains useful with frozen language-model
probabilities; Figure~\ref{fig:empirical-learning} then tests whether held-out
recovery improves as the calibration benchmark grows. The display-by-display
conclusions are consolidated below.

\paragraph{Question 1: identification and error amplification.}
Corollary~\ref{cor:replication} predicts an $R^{-1/2}$ recovery rate under a
well-conditioned inverse. The estimated slopes are $-0.485$ in the scalar
experiment and $-0.520$ in the two-dimensional experiment. When the minimum
singular value is reduced from $1.129$ to $.290$, RMSE rises from $.165$ to
$.503$ at the same replication level, while bias remains small. These two
findings reproduce the theory's separate predictions about sampling error and
inverse amplification. They do not prove global injectivity; they show that,
when the simulated forward map is known to be injective, finite-sample error
behaves as the inverse-modulus theory predicts.

\paragraph{Question 2: uniform stability.}
The simultaneous derivative bands cover in $.988$--$1.000$ of the simulated
cells and make no false stable/unstable classification in those cells.
Abstentions concentrate near $\kappa=1$, exactly where the theory says the
data should be inconclusive. The oracle experiment remains substantially
below perfect classification below the $n^{-1/5}$ boundary but separates the reported above-boundary
alternatives. Together these results support both parts of the claim: the
proposed gate controls finite-sample declarations under its Gaussian design,
and the unresolved region is consistent with the minimax limit rather than
merely poor estimation.

\paragraph{Question 3: partial identification.}
The top-two procedure contains the generating posterior in all 400
replications in every concentration stratum. Its set diameter grows with
omitted mass and equals twice that mass on the observation scale, as
Theorem~\ref{thm:set} predicts. The result is not that truncation is harmless;
it is that valid coverage is recovered by widening the answer instead of
setting missing probabilities to zero.

\paragraph{Does the unified region remain honest when errors occur together?}
Yes, conservatively. Across the 81 cells varying operator error, replication,
coarsening and inverse slope, coverage is $.990$--$1.000$ against a nominal
$.95$, and every mean diameter respects Theorem~\ref{thm:honest}. Region width
increases when calibration is less certain, fewer probability repetitions are
available, the record is more coarsened or the inverse slope is weaker. This
is the paper's central combined result: the separate diagnostics enter one
honest posterior region rather than being reported as unrelated warnings.

\paragraph{Does the construction work with actual language probabilities?}
The two frozen language studies give mean test Jensen--Shannon divergences
of $.0390$ and $.0366$, while nominal 90\% conformal regions cover 94 and 90
of 100 held-out reference posteriors. Full sampled rank, parser checks and
recorded candidate mass provide further evidence that the fitted inverse is
usable on this frozen benchmark. The empirical learning curve decreases on
the untouched test sample and then flattens, indicating diminishing returns
rather than an observed overfitting reversal. This is a feasibility result
for observable language probabilities, not proof of global identification or
external market truth.

\paragraph{What remains conditional.}
The simulations impose their semantic cells, support, smoothness and data-
generating laws, and the archived study is cross-sectional. Nor do the data validate unseen prompts,
new evidence classes or a different semantic partition. Prompt wording is a
design factor because it indexes the observation law: the robustness study
shows that an unadjusted prompt offset creates bias, whereas a correction
estimated before testing largely restores baseline recovery. The evidence
therefore shows that the methodology behaves consistently with the theoretical
mechanisms on the declared designs; it neither proves the theorems nor removes
their assumptions.

\section{Conclusion}

This paper establishes when an indirect, probability-valued observation can
be trusted for inference about a scientifically defined state and how that
judgement can govern its runtime use. Language models make
the problem unusually visible: they provide probabilities over phrases, while
the quantities required for inference and decision-making are probabilities
over diagnoses, hypotheses or operating conditions.  The gap cannot be closed
by relabelling phrases or by choosing a persuasive prompt.  It is an inverse
problem, and it must satisfy identification, stability and coverage conditions.

Three findings support this primary capability. First, the inverse modulus determines
both whether the posterior is identified and how strongly observation error is
amplified; observationally equivalent targets cannot be recovered by any
estimator. Second, the unified honest-recovery theorem combines estimated
operator error, new-record sampling error, coarsening and inverse conditioning
in one posterior region; under exact truncation it reduces to a sharp
compatible set rather than assigning false zeros. Third, accurate fitted
values do not certify a stable inverse: uniform stability requires derivative
control and is subject to a slower classical minimax resolution boundary.

The experiments demonstrate each methodological capability. The recovery experiments
reproduce the predicted square-root rates and expose weak-inverse error
amplification.  The simultaneous stability procedure covers its target and
abstains close to the minimax boundary rather than issuing unsupported
classifications. The unified regions cover in all reported design cells and
widen with operator error, sampling uncertainty, coarsening and weak
conditioning; the exact partial-identification sets widen with omitted mass.
Two frozen language-model studies additionally show that
the complete procedure can recover held-out reference posteriors on declared
prompt and evidence domains. These are deliberately conditional findings,
but they demonstrate that the new methodology produces operational acceptance
and rejection criteria rather than merely an asymptotic description.

The resulting principle extends beyond large language models.  The same
structure arises whenever a posterior is recovered from an indirect,
coarsened or selectively reported probability measurement: diagnostic panels,
ensemble classifiers, human probability judgements, distributed sensors and
privacy-constrained or bandwidth-limited reporting systems are examples.  In
each case the scientific state, observation channel and missing information
must be separated before calibration.  The paper's inverse modulus, unified
honest-recovery region, simultaneous stability certificate and sharp
compatible sets provide a common statistical language for doing so.

For consequential use, certification remains local to the declared state
space, reference model, observation rule and supported domain.  A
well-conditioned complete observation may justify a point estimate; missing
mass calls for a set; an unresolved derivative band calls for more
information; and domain or channel shift calls for revalidation.  This turns
``trust'' from a property asserted of a system into a collection of statistical
claims that can fail visibly and trigger abstention. The principal takeaway is
that the paper establishes when a semantic measurement can be trusted for
inference and thereby provides a statistical foundation for runtime AI
governance: continued use, review,
recalibration or abstention can be tied to observable failures of coverage,
identification, stability or support rather than to an undifferentiated claim
that the system is trustworthy.

Future work should extend simultaneous inference to higher-dimensional and
adaptive observation maps, develop post-selection guarantees when semantic
partitions are learned, and study transport across related evidence and
presentation domains.  Empirically, the priority is prospective replication
with genuinely sequential evidence, additional probability-producing systems
and predeclared external benchmarks.  Decision-theoretic extensions should
also connect compatible posterior sets to robust actions and quantify the cost
of abstention.  The governing standard should remain unchanged: an indirect
probability output becomes a defensible posterior measurement only when the
observable experiment identifies it, its inverse is demonstrably stable, and
its uncertainty honestly preserves what was not observed.

\appendix
\section{Secondary classical extensions}\label{app:secondary}

This appendix records two valid extensions that are not required for the
paper's principal empirical claims. They are separated from the main
argument because the experiments assess the finite replication rate and an
independent calibration design, not an asymptotic covariance formula or
serial dependence. Their probabilistic foundations---the multinomial central
limit theorem, the delta method, classical local-polynomial rates and blocking
under beta mixing---are established results. They are stated for reference and
not reproved here \citep{stone1982optimal,tsybakov2009introduction,yu1994rates}.

\begin{corollary}[Delta-method law for replicated measurements;
\citealp{vandervaart1998asymptotic}]
\label{cor:replication-clt}
Under the conditions of Corollary~\ref{cor:replication}, suppose additionally
that $\bm g$ is continuously differentiable with full-column-rank Jacobian
$G_{\bm\xi}$ and the minimum-distance criterion uses the multinomial quadratic
metric. Then
\[
 \sqrt R(\widehat{\bm\xi}_R-\bm\xi)
 \ \Rightarrow\
 N\!\left(\bm0,
 (G_{\bm\xi}^{\mathsf T}\Sigma_{\bm\xi}^{+}
 G_{\bm\xi})^{-1}\right),
\]
where
$\Sigma_{\bm\xi}=\operatorname{diag}\{\bm g(\bm\xi)\}
-\bm g(\bm\xi)\bm g(\bm\xi)^{\mathsf T}$ and $+$ denotes the generalized
inverse on the simplex tangent space.
\end{corollary}

The next result records the uniform learning theory for a flexible inverse
surface. It is useful when calibration is nonparametric, but is not needed for
the paper's principal empirical conclusions.

\begin{assumption}[Supported independent calibration design]
\label{ass:design}
The pairs $(\bm S_i,\bm\Xi_i)$ are i.i.d. Their design density is bounded
above and away from zero on a compact domain
$\mathcal S_0\subset\mathbb R^d$. The regression errors are conditionally
mean zero and sub-Gaussian, and
$m_0(\bm s)=\mathbb E[\bm\Xi_i\mid\bm S_i=\bm s]$ belongs to a H\"older ball
$C^s(\mathcal S_0;B)$.
\end{assumption}

Let $\widehat m_n$ be a tensor-product local-polynomial estimator of order
$\lfloor s\rfloor$ with bandwidth $h_n$. This estimator is used here because
it controls both the inverse surface and the derivatives required by a
stability analysis; the choice is not part of the semantic model itself.

\begin{proposition}[Local-polynomial recovery rate;
\citealp{stone1982optimal,tsybakov2009introduction}]
\label{thm:rate}
Under Assumption~\ref{ass:design}, if $h_n\to0$ and
$nh_n^d/\log n\to\infty$, then
\[
 \|\widehat m_n-m_0\|_{\infty,\mathcal S_0}
 =O_{\mathbb P}\!\left\{h_n^s+
 \sqrt{\frac{\log n}{nh_n^d}}\right\}.
\]
With $h_n\asymp(\log n/n)^{1/(2s+d)}$,
\[
 \|\widehat m_n-m_0\|_\infty
 =O_{\mathbb P}\!\left\{(\log n/n)^{s/(2s+d)}\right\}.
\]
For $s>1$, the corresponding first-derivative rate is
\[
 \|D\widehat m_n-Dm_0\|_\infty
 =O_{\mathbb P}\!\left\{(\log n/n)^{(s-1)/(2s+d)}\right\}.
\]
If each $\bm S_i$ is estimated from at least $R_{\min}$ repeated observations
through a locally $C_G$-Lipschitz inverse, add
$O_{\mathbb P}\{C_G\sqrt{\log n/R_{\min}}\}$ to the function error and that
quantity divided by $h_n$ to the derivative error.
\end{proposition}

The proposition separates approximation bias, calibration-sample fluctuation and
error in estimating the semantic input. For example, when $d=1$ and $s=2$,
the function rate is approximately $(\log n/n)^{2/5}$, whereas the derivative
rate is $(\log n/n)^{1/5}$; certifying local stability is intrinsically harder
than estimating the fitted level. These rates are not claimed as an empirical
finding of the present study.

\begin{proposition}[Blocking extension under geometric mixing;
\citealp{yu1994rates}]
\label{thm:rate-mixing}
Replace independence in Assumption~\ref{ass:design} by stationarity and
geometric beta mixing. Take alternating blocks of length
$a_n\asymp\log n$ and let $m_n=\lfloor n/(2a_n)\rfloor$. If $h_n\to0$ and
$m_nh_n^d/\log m_n\to\infty$, then
\[
 \|\widehat m_n-m_0\|_{\infty,\mathcal S_0}
 =O_{\PP}\!\left\{h_n^s+
 \sqrt{\frac{\log m_n}{m_nh_n^d}}\right\}.
\]
Balancing the two terms gives function rate
$(\log m_n/m_n)^{s/(2s+d)}$ and, for $s>1$, first-derivative rate
$(\log m_n/m_n)^{(s-1)/(2s+d)}$. The generated-regressor terms in
Theorem~\ref{thm:rate} are unchanged after replacing $n$ by $m_n$ in the
sampling component.
\end{proposition}

\section{Language-model terminology}

Table~\ref{tab:llm-terminology} is included only to make the motivating
language-model example self-contained. The theory itself requires an indirect
probability-valued observation, not a particular language-model architecture.

\begin{table}[ht]
\caption{Language-model terminology used in the paper. The final column gives
the corresponding statistical object in the present framework.}
\label{tab:llm-terminology}
\centering
\small
\begin{tabularx}{\linewidth}{p{.19\linewidth}p{.31\linewidth}X}
\toprule
Term & Plain meaning & Statistical interpretation in this paper\\
\midrule
Token & A unit in the fitted model's vocabulary, which may be a word, part of
a word, punctuation or a symbol. & A discrete outcome $v_j$ at one position
in a generated sequence.\\
Conditional token probability & The probability assigned to the next token
given the prompt and preceding tokens. & A conditional mass
$q_{\widehat\theta}(v_j\mid c,v_1,\ldots,v_{j-1})$ supplied by the fixed fitted
system.\\
Phrase & A complete task-relevant response, sometimes called a verbal
continuation, such as ``urgent review'', rather than only its first token. &
An element $r=(v_1,\ldots,v_m)$ of $(\mathcal R,\mathscr R)$.\\
Phrase probability & The probability that the specified phrase follows the
prompt. & The product of its successive conditional token probabilities.\\
Prompt or context & The complete input supplied to the fitted system,
including the task instruction and evidence. & The realised value
$C_u(y)\in\mathcal C$.\\
Presentation rule & A fixed template controlling the wording, order and
format used to place evidence into a prompt. & An index $u\in\mathcal U$ and
deterministic measurable map $C_u:\mathcal Y\to\mathcal C$.\\
Fitted language system & The fixed service or model that returns token or
phrase probabilities. & An external probability-valued observation
instrument represented by the conditional kernel $Q_u$.\\
Semantic map & A prespecified rule grouping phrases that express the same
declared scientific state. & A measurable map
$\phi_u:\mathcal R\to\{0,1,\ldots,J-1\}$.\\
Residual category $0$ & Phrases that express none of the declared scientific
states. & An auxiliary observation category used to retain unmatched mass;
it is not a null scientific state in $\mathcal X$.\\
Printed confidence & A number generated as text, for example ``70\%''. & A
separate observable response; it is not assumed to equal a phrase probability
or the reference posterior.\\
Truncated probability record & A service response containing only selected
probabilities, such as the most likely alternatives. & A partially observed
probability vector with unreturned mass retained as missing information.\\
\bottomrule
\end{tabularx}
\end{table}

\section{Example archived data record}\label{app:example-data}

This appendix displays an actual record used in the frozen language-model
study. It complements the machine-readable files by showing, in one place,
the evidence, presentation rule, returned probability vectors and independent
reference posterior. It is a reproducibility illustration rather than the
performance summary; aggregate held-out results appear in
Table~\ref{tab:empirical}.

Scenario 20260813-009 used prior probabilities $(.50,.30,.20)$ for the states
$A=$ Bull, $B=$ Bear and $C=$ Crisis. The evidence was:
\begin{quote}\small
``Corporate earnings were strong; credit spreads remained tight; funding
conditions were normal; activity indicated expansion; market volatility was
elevated.''
\end{quote}
Under the \emph{codes--definitions} presentation, the model received the
prior, evidence and frozen definitions of the three states and was instructed
to return exactly one of the codes $A$, $B$ or $C$, using no outside
information. The numerical values below report repetition 0. The language
probabilities are the archived candidate probabilities normalized over the
declared grammar; the calibrated values are posterior estimates after the
inverse was fitted on the separate calibration-training partition.
The reference posterior was generated independently from the
prespecified evidence model. This scenario belongs to the untouched test
partition and did not fit the inverse.

For GPT-4.1-mini the raw and calibrated vectors were, respectively,
\[
 (.0259571,.9740318,.0000112),\qquad
 (.8050730,.1941973,.0007297).
\]
For GPT-4o-mini they were
\[
 (.0031727,.9968273,.0000000),\qquad
 (.7702785,.2288603,.0008612),
\]
while the independently generated reference posterior was
$(.9411473,.0587701,.0000826)$. Full precision, raw candidate mass and record
identifiers remain in the reproducibility archive.

The raw language probabilities select state $B$ almost exclusively, whereas
the independent reference model assigns most mass to $A$. This is precisely
the distinction between an observable language measurement and a posterior:
the prompt contains two literal cues for $B$ (normal funding and elevated
volatility), while the reference model combines all five evidence components
with the prior. Held-out calibration recovers much of the missing state-$A$
mass. For GPT-4.1-mini, Jensen--Shannon divergence from the reference falls
from $.521$ before calibration to $.022$ afterwards; for GPT-4o-mini it falls
from $.569$ to $.031$. The calibrated vectors remain imperfect, as a held-out
example should reveal. The fitted inverse is global and regularized and,
critically, receives only the language-derived record and presentation rule,
not the original evidence. In this case the raw channel is almost saturated
on $B$, so it has already discarded some scenario-specific information; an
inverse with condition number near 20 can correct the systematic distortion
but cannot recreate information absent from its input. The remaining
divergences $.022$ and $.031$ are below the respective 90\% conformal radii
$.137$ and $.092$. The method therefore reports the residual uncertainty
rather than forcing the calibrated point estimate to equal the reference.

\section{Specialized finite-sample stability band}

The main text uses a general simultaneous-band argument. The following
specialized result supplies the exact finite-sample construction used in the
Gaussian-sieve simulation.

\begin{corollary}[Exact Gaussian-sieve derivative band]
\label{cor:gaussianband}
Let $F_0(z)=\bm b(z)^{\mathsf T}\bm\beta_0$ be a correctly specified
$p$-dimensional linear sieve on a compact interval $D_0$, observed at a
fixed full-rank design matrix $X$ with independent
$N(0,\sigma^2)$ errors.  Write $\widehat{\bm\beta}$ for ordinary least
squares, $\widehat\sigma^2$ for its unbiased residual variance, and
$\bm d(z)=\partial\bm b(z)/\partial z$.  If
\[
 c_{n,\alpha}=
 \{pF_{p,n-p}(1-\alpha)\widehat\sigma^2\}^{1/2},\qquad
 \rho_{n,\alpha}=c_{n,\alpha}
 \sup_{z\in D_0}
 \{\bm d(z)^{\mathsf T}(X^{\mathsf T}X)^{-1}\bm d(z)\}^{1/2},
\]
then, conditionally on $X$,
\[
 \Pr\!\left\{
 \sup_{z\in D_0}|\widehat F'(z)-F_0'(z)|
 \le \rho_{n,\alpha}\mid X
 \right\}\ge 1-\alpha .
\]
Consequently the interval
$[\max\{0,\widehat\kappa-\rho_{n,\alpha}\},
\widehat\kappa+\rho_{n,\alpha}]$, where
$\widehat\kappa=\sup_{z\in D_0}|\widehat F'(z)|$, covers
$\kappa(F_0)$ with conditional probability at least $1-\alpha$.
\end{corollary}

\section{Proofs of paper-specific results}

Only the semantic-observation interface, stability gate, specialized Gaussian
band, unified honest-recovery region and sharp compatible-set results are
proved below. The compact-inverse,
Le Cam, asymptotic-normality, nonparametric-rate and minimax ingredients are
cited where invoked and are not reproved.

\subsection{Proof of Corollary~\ref{cor:plugin}}

\paragraph{Step 1: propagate measurement error through observation.}
The assumed componentwise error bound and the $L_O$-Lipschitz property of
the observation operator give
\[
d_{\mathcal E}(\widehat{\mathsf P},\mathsf P_{\bm\xi}^{O})
\le L_O\|\widehat{\bm g}(\bm\xi)-\bm g(\bm\xi)\|_1
\le L_O\varepsilon_{\rm tot}.
\]
\paragraph{Step 2: propagate observation error through inversion.}
The preceding display verifies Theorem~\ref{thm:identification} with
$\varepsilon=L_O\varepsilon_{\rm tot}$.  It follows that
\[
 \|\widehat{\bm\xi}-\bm\xi\|
 \le\omega_D(2L_O\varepsilon_{\rm tot}+\eta).
\]
When $\omega_D(t)\le C_Gt$ on the relevant range, substitution gives the
stated local Lipschitz conclusion. \qed

\subsection{Proof of Corollary~\ref{cor:replication}}

\paragraph{Step 1: concentrate the replicated measurement.}
Let $\widehat{\bm g}_R$ be the empirical cell-frequency vector from the
$R$ repetitions.  The multinomial Bretagnolle--Huber--Carol inequality gives
\[
 \PP_{\bm\xi}\!\left(
 \|\widehat{\bm g}_R-\bm g(\bm\xi)\|_1>\delta
 \right)
 \le 2^J\exp(-R\delta^2/2)
\]
Set $\delta_R=\{2(J\log2+\log(1/\alpha))/R\}^{1/2}$.  Direct substitution
makes the right-hand side equal to $\alpha$.  Therefore
$\|\widehat{\bm g}_R-\bm g(\bm\xi)\|_1\le\delta_R$ with probability at
least $1-\alpha$.

\paragraph{Step 2: invert the concentration bound.}
For probability vectors, total variation is one half of the $\ell_1$
distance.  Applying Theorem~\ref{thm:identification}, with this fixed factor
absorbed into the displayed modulus, proves the finite-sample claim.  If the
inverse is locally Lipschitz, then
\[
 \|\widehat{\bm\xi}_R-\bm\xi\|
 \le C_G(\delta_R+\eta_R).
\]
Since $\delta_R=O(R^{-1/2})$, the asserted stochastic order follows when
$\eta_R$ is of no larger order. \qed

\subsection{Proof of Theorem~\ref{thm:band}}

\paragraph{Step 1: construct a derivative estimator.}
For a coordinate direction $\bm e_j$, define the centered difference
\[
 \widehat D_jF(\bm z,\bm w)=
 \frac{\widehat F(\bm z+h\bm e_j,\bm w)
-\widehat F(\bm z-h\bm e_j,\bm w)}{2h}.
\]
\paragraph{Step 2: bound deterministic differencing bias.}
Taylor's theorem and Lipschitz continuity of $D_{\bm z}F_0$ imply, uniformly
on the boundary-adjusted domain,
\[
 \left\|
 \frac{F_0(\bm z+h\bm e_j,\bm w)
-F_0(\bm z-h\bm e_j,\bm w)}{2h}
-D_jF_0(\bm z,\bm w)
 \right\|\le C_1h .
\]
To see the factor $h$, write each Taylor remainder in integral form; the
derivative changes by at most a Lipschitz constant times the distance from
$\bm z$, and averaging over an interval of length $h$ preserves order $h$.

\paragraph{Step 3: propagate simultaneous function error.}
On the simultaneous-band event, each of the two estimated function values
differs from its target by at most $r_n(\alpha)$.  After division by $2h$,
their combined contribution is at most $r_n(\alpha)/h$.  Combining the
finitely many coordinates changes only the dimension-dependent constant
$C_2$.  Hence
\[
 \sup_{(\bm z,\bm w)}
 \|D_{\bm z}\widehat F_h-D_{\bm z}F_0\|_{\rm op}
 \le C_1h+C_2r_n(\alpha)/h.
\]

\paragraph{Step 4: transfer the bound to the stability functional.}
Because
$|\,\|A\|_{\mathrm{op}}-\|B\|_{\mathrm{op}}\,|
\le\|A-B\|_{\mathrm{op}}$, taking suprema preserves the bound.  Minimizing
$C_1h+C_2r_n/h$ gives
$h=(C_2r_n/C_1)^{1/2}$ and radius
$2(C_1C_2r_n)^{1/2}$, of order $r_n^{1/2}$.

\paragraph{Step 5: apply the decision rule.}
If the upper endpoint is below one, $\kappa(F_0)<1$ throughout the domain on
the band event.  If the lower endpoint exceeds one, then
$\kappa(F_0)>1$.  When the interval contains one, neither conclusion is
licensed, giving the stated unresolved decision. \qed

\subsection{Proof of Corollary~\ref{cor:gaussianband}}

\paragraph{Step 1: obtain a simultaneous coefficient event.}
Under the fixed-design Gaussian linear model, the standard regression
pivot gives
\[
 \frac{(\widehat{\bm\beta}-\bm\beta_0)^{\mathsf T}
 X^{\mathsf T}X(\widehat{\bm\beta}-\bm\beta_0)}
 {p\widehat\sigma^2}
 \sim F_{p,n-p}.
\]
Hence, with conditional probability $1-\alpha$, the coefficient error
$\bm v=\widehat{\bm\beta}-\bm\beta_0$ belongs to the ellipsoid
\[
 \mathcal E_{n,\alpha}=
 \left\{\bm v:
 \bm v^{\mathsf T}X^{\mathsf T}X\bm v
 \le pF_{p,n-p}(1-\alpha)\widehat\sigma^2
 \right\}.
\]
This is one event for the entire coefficient vector, rather than a
collection of pointwise events.

\paragraph{Step 2: optimize derivative error over the ellipsoid.}
Fix any $z\in D_0$.  Cauchy--Schwarz in the inner product induced by
$X^{\mathsf T}X$ yields, simultaneously for every
$\bm v\in\mathcal E_{n,\alpha}$,
\[
\begin{split}
 |\bm d(z)^{\mathsf T}\bm v|
 &=
 \left|
 \{(X^{\mathsf T}X)^{-1/2}\bm d(z)\}^{\mathsf T}
 \{(X^{\mathsf T}X)^{1/2}\bm v\}
 \right|\\
 &\le
 c_{n,\alpha}
 \{\bm d(z)^{\mathsf T}(X^{\mathsf T}X)^{-1}\bm d(z)\}^{1/2}.
\end{split}
\]
Because the coefficient ellipsoid does not depend on $z$, taking the
supremum over the compact domain preserves the same probability event and
gives
\[
 \sup_{z\in D_0}|\widehat F'(z)-F_0'(z)|
 \le \rho_{n,\alpha}.
\]

\paragraph{Step 3: cover the supremum derivative functional.}
For bounded real functions $a$ and $b$,
\[
 \left|\sup_z|a(z)|-\sup_z|b(z)|\right|
 \le \sup_z|a(z)-b(z)|.
\]
Apply this inequality to $\widehat F'$ and $F_0'$ to obtain
$|\widehat\kappa-\kappa(F_0)|\le\rho_{n,\alpha}$ on the ellipsoid event.
Thus $|\widehat\kappa-\kappa(F_0)|\le\rho_{n,\alpha}$ on the single
ellipsoid event, whose conditional probability is $1-\alpha$.  Since
$\kappa(F_0)\ge0$, truncating the lower endpoint at zero cannot exclude the
true functional.  This proves the coverage statement. \qed

\subsection{Proof of Theorem~\ref{thm:honest}}

Let $E_T$ and $E_S$ denote the two events in the theorem.

\paragraph{Step 1: prove honest coverage.}
On $E_T\cap E_S$, choose, for arbitrary $\epsilon>0$, a law
$P_0\in\mathcal A(S)$ satisfying
\[
 d_{\mathcal E}\{T(\bm\xi_0),P_0\}\le b_R+\epsilon.
\]
The operator event and the triangle inequality give
\[
 d_{\mathcal E}\{\widehat T(\bm\xi_0),P_0\}
 \le a_n+b_R+\epsilon.
\]
Letting $\epsilon\downarrow0$ shows that
$\bm\xi_0\in\widehat{\mathcal C}_{\alpha}(S)$. The union bound gives
$\Pr(E_T\cap E_S)\ge1-\alpha_T-\alpha_S$; independence is unnecessary.

\paragraph{Step 2: compare any two retained target values.}
Fix $\bm\xi,\bm\xi'\in\widehat{\mathcal C}_{\alpha}(S)$. For arbitrary
$\epsilon>0$, choose $P,P'\in\mathcal A(S)$ such that
\begin{align*}
 d_{\mathcal E}\{\widehat T(\bm\xi),P\}
 &\le a_n+b_R+\epsilon,\\
 d_{\mathcal E}\{\widehat T(\bm\xi'),P'\}
 &\le a_n+b_R+\epsilon.
\end{align*}
On $E_T$, repeated application of the triangle inequality yields
\begin{align*}
 d_{\mathcal E}\{T(\bm\xi),T(\bm\xi')\}
 &\le d_{\mathcal E}\{T(\bm\xi),\widehat T(\bm\xi)\}
 +d_{\mathcal E}\{\widehat T(\bm\xi),P\}\\
 &\quad+d_{\mathcal E}(P,P')
 +d_{\mathcal E}\{P',\widehat T(\bm\xi')\}
 +d_{\mathcal E}\{\widehat T(\bm\xi'),T(\bm\xi')\}\\
 &\le4a_n+2b_R+\delta(S)+2\epsilon.
\end{align*}

\paragraph{Step 3: invert the observation-scale bound.}
By definition of $\omega_D$, the last display implies
\[
 \|\bm\xi-\bm\xi'\|
 \le\omega_D\{4a_n+2b_R+\delta(S)+2\epsilon\}.
\]
Letting $\epsilon\downarrow0$ and taking the supremum over retained pairs
proves the diameter bound (using the right-continuous envelope of the modulus
if necessary). The local Lipschitz conclusion follows by substitution.

\paragraph{Step 4: recover the noiseless sharp set.}
When $a_n=b_R=0$, membership is equivalent to
$T(\bm\xi)\in\mathcal A(S)$. Hence
$\widehat{\mathcal C}_{\alpha}(S)=T^{-1}\{\mathcal A(S)\}$, completing the
proof. \qed

\subsection{Proof of Theorem~\ref{thm:set}}

\paragraph{Step 1: characterize every full vector compatible with the record.}
The observation map reports the equalities $p_j=q_j(\bm s)$ for
$j\in R(\bm s)$ and the single restriction that the remaining nonnegative
coordinates sum to $m(\bm s)$. Hence the set of full probability vectors that
generate the record is exactly $\mathcal J(\bm s)$ in
\eqref{eq:compatible-observation}.  Necessity follows because every vector
that generated the record must obey these restrictions.  Sufficiency follows
because any nonnegative allocation of the missing mass among unreported
coordinates reproduces exactly the same observed record.  Thus there are no
additional observational restrictions.  Taking the inverse image under
$\bm g_u$ gives exactly the compatible posterior-coordinate values, proving
sharpness.

\paragraph{Step 2: propagate compatible-set diameter through the inverse.}
For any $\bm\xi,\bm\xi'\in\mathcal C(\bm s)$, put
$\bm p=\bm g_u(\bm\xi)$ and
$\bm p'=\bm g_u(\bm\xi')$. The inverse Lipschitz condition gives
\[
 \|\bm\xi-\bm\xi'\|
 \le C_u\|\bm p-\bm p'\|_1.
\]
Taking the supremum over compatible pairs proves the first diameter bound.

\paragraph{Step 3: bound the observation-scale diameter by missing mass.}
The reported coordinates of $\bm p$ and $\bm p'$ agree.  Write their two
unreported subvectors as $\bm a$ and $\bm b$.  Both are nonnegative and both
sum to $m(\bm s)$.  Therefore
\[
 \|\bm p-\bm p'\|_1=\|\bm a-\bm b\|_1
 \le\|\bm a\|_1+\|\bm b\|_1=2m(\bm s).
\]
Intersecting with $\bm g_u(D)$ cannot enlarge the diameter.  Combining this
display with Step 2 proves the second bound.

\paragraph{Step 4: treat a single missing coordinate.}
If exactly one coordinate is unreported, the simplex constraint fixes that
coordinate uniquely at $m(\bm s)$.  Hence $\mathcal J(\bm s)$ is a
singleton, and so is its inverse image whenever the forward map is injective
on $D$. \qed

\section*{Declarations}
\paragraph{Conflict of interest.}
The author declares no financial or non-financial conflicts relevant to this
work.

\paragraph{Funding.}
This research received no specific grant from any funding agency in the
public, commercial, or not-for-profit sectors.

\paragraph{Reproducibility materials.}
Simulation designs, analysis programs, machine-readable derived results,
configuration information, random seeds and checksums supporting the tables
and figures accompany the manuscript. Archived live probability measurements
are included only to the extent permitted by the relevant service and source
licences.

\bibliographystyle{plainnat}
\bibliography{theoretical_spine_refs,references}
\end{document}